\documentclass[12pt, letterpaper]{article}

\usepackage{lineno,hyperref}

\usepackage{subcaption}
\usepackage{algorithm}
\usepackage{algpseudocode} % use algorithmics package
\usepackage[T1]{fontenc} % for \k
\usepackage{authblk}
\usepackage{graphicx}

\newcommand{\Var}{\mathrm{Var}} % define variance

\title{The Impact of Position Errors on Crowd Simulation}

\author[1]{Lei Zhang\thanks{ Email address: \href{mailto:leizhang@ryerson.ca}{leizhang@ryerson.ca}}}
\author[2]{Diego Lai}
\author[1]{Andriy V. Miranskyy\thanks{Email address: \href{mailto:avm@ryerson.ca}{avm@ryerson.ca}}}
\affil[1]{Department of Computer Science, Ryerson University, Toronto, Canada}
\affil[2]{Laipac Technology Inc., Toronto, Canada}

\date{}

\begin{document}
\maketitle

\begin{abstract}
In large crowd events, there is always a potential possibility that a stampede accident will occur. The accident may cause injuries or even death. Approaches for controlling crowd flows and predicting dangerous congestion spots would be a boon to on-site authorities to manage the crowd and to prevent fatal accidents. One of the most popular approaches is real-time crowd simulation based on position data from personal Global Positioning System (GPS) devices. However, the accuracy of spatial data varies for different GPS devices, and it is also affected by an environment in which an event takes place. In this paper, we would like to assess the effect of position errors on stampede prediction. We propose an Automatic Real-time dEtection of Stampedes (ARES) method to predict stampedes for large events. We implement three different stampede assessment methods in Menge framework and incorporate position errors. Our analysis suggests that the probability of simulated stampede changes significantly with the increase of the magnitude of position errors, which cannot be eliminated entirely with the help of classic techniques, such as the Kalman filter. Thus, it is our position that novel stampede assessment methods should be developed, focusing on the detection of position noise and the elimination of its effect. 
\end{abstract}

\section{Introduction}\label{sec:introduction}

In recent decades, the number of events where a stampede has occurred is increasing, along with the number of people involved in such accidents. In 2014, 36 people were killed and 47 others were injured in a crush at New Year's Eve celebrations in Shanghai, China~\cite{online:shanghai}; in 2015, at least 2,177 people were crushed to death and 934 were injured at the annual Hajj in Saudi Arabia~\cite{online:hajj}; in 2017, more than 1,500 soccer fans were injured in stampede in Turin, Italy~\cite{online:turin}; in the same year, at least 22 people were killed and hundreds were injured in a stampede at Elphinstone road station in Mumbai, India~\cite{online:mumbai}. 

To prevent stampede accidents, the Saudi Ministry of Hajj has already begun considering plans to give each pilgrim an electronic bracelet with individual identification and Global Positioning System (GPS)~\cite{online:ebracelet}. Moreover, the percentage of the population owning smartphones (having built-in GPS) is increasing. The recent reports show that ten countries now have smartphone penetration\footnote{That is the percentage of the people in the population that own a smartphone.} greater than $70\%$~\cite{online:smartphone_penetration}, with the numbers growing each year~\cite{online:smartphone_worldwide}. This will provide a technological foundation to enable the gathering of locations of most of the people in the near future, even if no specialized device (such as the electronic bracelet mentioned above) is present.

In recent years, many researches are leveraging data from GPS handheld devices to prevent stampedes by providing early warning to authorities~\cite{wirz2012inferring, franke2015smart, zhou2017early}. However, all GPS devices suffer from position errors. The main GPS error source is due to the satellite synchronization~\cite{liu2012energy}. Other errors arise because of atmospheric disturbances that distort the signals before they reach a receiver. Reflections from buildings and other large, solid objects can lead to GPS accuracy problems too. To the best of our knowledge, the literature is lacking the analysis of the impact of GPS noise on the prediction of stampedes.

In this paper, we focus on quantifying the impact of GPS measurement errors on stampede probabilities. The main contributions of this paper include: (1) presentation of a stampede prediction solution by the incorporation of three different stampede assessment methods into a crowd simulator; (2) introduction of a noise modeling technique in crowd simulation and evaluation of the probability of simulated stampede with various position errors; and (3) analysis of the impact of position errors on simulated stampede probabilities.

The remainder of this paper is constructed as follows: Section~\ref{sec:related} reviews the related work in crowd management, stampede assessment methods, and crowd simulation. Section~\ref{sec:method} describes our proposed solution, namely, how we simulate stampede accidents and measure the position errors. Section~\ref{sec:experiments} covers experimental results and analysis. Finally, Section~\ref{sec:conclusions} summarizes our findings.

\section{Related Work}\label{sec:related}
% GPS-based
Crowd management based on data from GPS handheld devices (e.g., smartphones, smartwatches, and personal fitness trackers) has been widely studied. Wirz et al.~\cite{wirz2012inferring} adopted the crowd pressure technique to visualize this information as heat maps, offered a global view of the crowd situation, and assessed different crowd conditions instantaneously throughout an event. Franke et al.~\cite{franke2015smart} implemented a smartphone based crowd management system, which also uses a heat map representation of the crowd state and its evolution. Zhou et al.~\cite{zhou2017early} proposed a solution based on Baidu map and developed a prediction model to perceive the crowd anomaly and to assess the risk of the crowd event. 

% visualized stampede models
Besides GPS-based solutions, vision-based systems have also been developed to detect congestion spots and pedestrian behaviors. The accuracy of vision-based systems can be affected by three factors: camera coverage, camera resolutions, and level of illumination~\cite{zhou2017early, krausz2012loveparade}. The existing computer vision work focuses on real-time detection of the stampedes~\cite{junior2010crowd, marana1999estimating, kurilkin2016comparison, mousavi2015analyzing, zhou2016spatial, sabokrou2017deep, sabokrou2018deep} rather than proactive prediction, making it complementary to our goal. While the field of computer vision is rapidly evolving, modern computer vision methods have not perfected crowd analysis yet. For example, they have difficulty tracking individuals in the crowd~\cite{junior2010crowd, marana1999estimating}. The training of a computer vision model for each camera has to be individualized~\cite{kurilkin2016comparison}. The accuracy of the detection of anomalies fluctuates for various methods, e.g., going to as low as 57\%~\cite{mousavi2015analyzing}, which may negatively affect the accuracy of stampede analysis. Last but not least, vision-based detection techniques face privacy and security related issues~\cite{sharma2016review} (which are conceptually similar to those based on GPS trackers even though a different technology is leveraged).

Recent research of anomaly detection in crowd focuses on developing deep learning-based approaches, e.g., Convolutional Neural Networks (CNNs), combined with video analysis~\cite{zhou2016spatial, sabokrou2017deep, sabokrou2018deep}. However, they have not yet been tested on high-density scenarios, which result in stampedes; so far the performance of CNNs was benchmarked against the UCSD dataset~\cite{online:ucsd}, which focuses on sparse crowds with non-pedestrian anomalous objects (e.g., skaters, biker, carts, and wheelchairs) and anomalous pedestrian patterns (e.g., a pedestrian walking in opposite direction to other pedestrians). Conceptually, one may create a hybrid stampede detection system that would leverage GPS trackers and video data feeds synergistically, improving the accuracy of the stampede analysis~\cite{bostanci2018sensor}. 

% crowd simulation
Agent-based pedestrian crowd simulation is a well-known solution for crowd management. Thus, a variety of simulation frameworks were proposed. Almeida et al.~\cite{almeida2011crowd} proposed a multi-agent framework to simulate emergency evacuation scenarios. Curtis et al.~\cite{curtis2016menge} introduced an open-source, cross-platform and agent-based crowd simulation framework---Menge. Mahmood et al.~\cite{mahmood2017analyzing} proposed an agent-based crowd simulation and analysis framework and used a case study of Hajj as an example for the assessment of crowd evacuation strategies.

% stampede models
To analyze the risk of stampedes, a number of stampede measuring techniques were presented in the literature. Helbing et al.~\cite{helbing2007dynamics}, Johansson et al.~\cite{johansson2008crowd} derived crowd pressure formulas (by analyzing video recordings of the crowd disaster during the Hajj in 2006) and discovered crowd turbulence phenomenon, in which physical contacts among people are transferred and accumulated. People try to escape the crowd, which causes crowd panic. Such phenomenon may trigger crowd disasters, such as stampedes. Lee and Hughes~\cite{lee2006prediction} developed a crowd pressure measuring technique using the standard forward-backward auto-regressive modeling approach. Elliott and Smith~\cite{elliott1993football} examined some sporting disasters and found the relationship between the accidents and the inter-person forces. Helbing et al.~\cite{helbing2002simulation} introduced social and physical forces among pedestrians and then treated each pedestrian as a particle abiding Newton's laws. Lee and Hughes~\cite{lee2005exploring} studied the relationship between crushing situation and people density, which gives the typical threshold of people density of different nationalities. However, none of these work has taken into account the impact of GPS noise on the prediction of stampedes.

% GPS accuracy 
Recent research has attempted to combine data from GPS and external sensors to improve the tracking accuracy. Groves et al.~\cite{groves2014four} developed a positioning technique by integrating GPS and Wi-Fi data (to improve positioning indoors) and accelerometers (to distinguish pedestrian behavior). Corbetta et al. proposed an approach to utilize overhead Microsoft Kinect\texttrademark depth sensors placed in a walkway to provide reliable automatic pedestrian positioning for tracking~\cite{corbetta2016continuous}. Suzuki et al.~\cite{suzuki2011high} proposed to use infrared cameras to eliminate the multi-path errors which are caused by different positioning systems. Assisted-GPS (A-GPS) technique is also widely used to improve the performance of GPS-enabled smartphones~\cite{van2009gps}. In general, combining GPS, WiFi, and sensors data improves tracking accuracy indoors (see~\cite{stojanovic2014indoor} for review). Besides employing external sensors, various algorithms were developed to mitigate GPS measurement noise, e.g., the Kalman filter~\cite{eliasson2014kalman, yamaguchi2006gps} and the autoregressive moving average filter~\cite{li2000gps}. 

As a rule of thumb, the accuracy of GPS can be hugely affected by the environment. The U.S. government claimed that GPS-enabled smartphones are typically accurate to within 4.9 m; however, their accuracy worsens near buildings or other obstacles~\cite{online:gps}. Zandbergen and Barbeau's research showed that the horizontal error of GPS-enabled smartphones ranges from 5.0 m to 8.5 m~\cite{zandbergen2011positional}. In 2015, Garnett and Stewart tested two personal GPS devices and two GPS-enabled mobile devices, and they found that the mean relative accuracy ranges from 3.65 m to 6.50 m~\cite{garnett2015comparison}.

\section{The ARES method}\label{sec:method}

In this paper, we propose a simulation-based method to predict a stampede during large crowd events. The name of the method is ARES, which stands for Automatic Real-time dEtection of Stampedes. In ARES, we assume that the venue map is provided before the event, and we can collect all pedestrian locations via personal GPS devices during the event. We call pedestrians with wearable GPS devices \textit{agents}. The word agent is used interchangeably with pedestrian in this paper. The method employs simulation to monitor pedestrians' behavior and measure possible stampede accidents. In case of a potential stampede, on-site authorities will be notified by an alert message, so that the authorities can intervene proactively and prevent the stampede from occurring. 

The key idea of ARES is to run the simulation and estimate the probability of a future stampede ahead of time. In reality, agent locations are collected in a discrete time manner to preserve GPS power. The interval between two GPS data collection points is denoted by $T$. During this time interval, GPS locations are unknown. Thus, we use the latest GPS locations as starting positions and simulate agents' trajectories for the interval $T$. To mitigate the uncertainty, we employ Monte Carlo method in simulations to calculate the stampede probabilities. We run simulations in parallel on a computing platform fast enough (e.g., a computer cluster) to ensure simulation completion time $K$ is less than $T$. Once the stampede probability is calculated, the result (including position information) is sent to the authorities, if necessary. 

\begin{figure}[!ht]
	\centering
	\includegraphics[width=0.9\columnwidth]{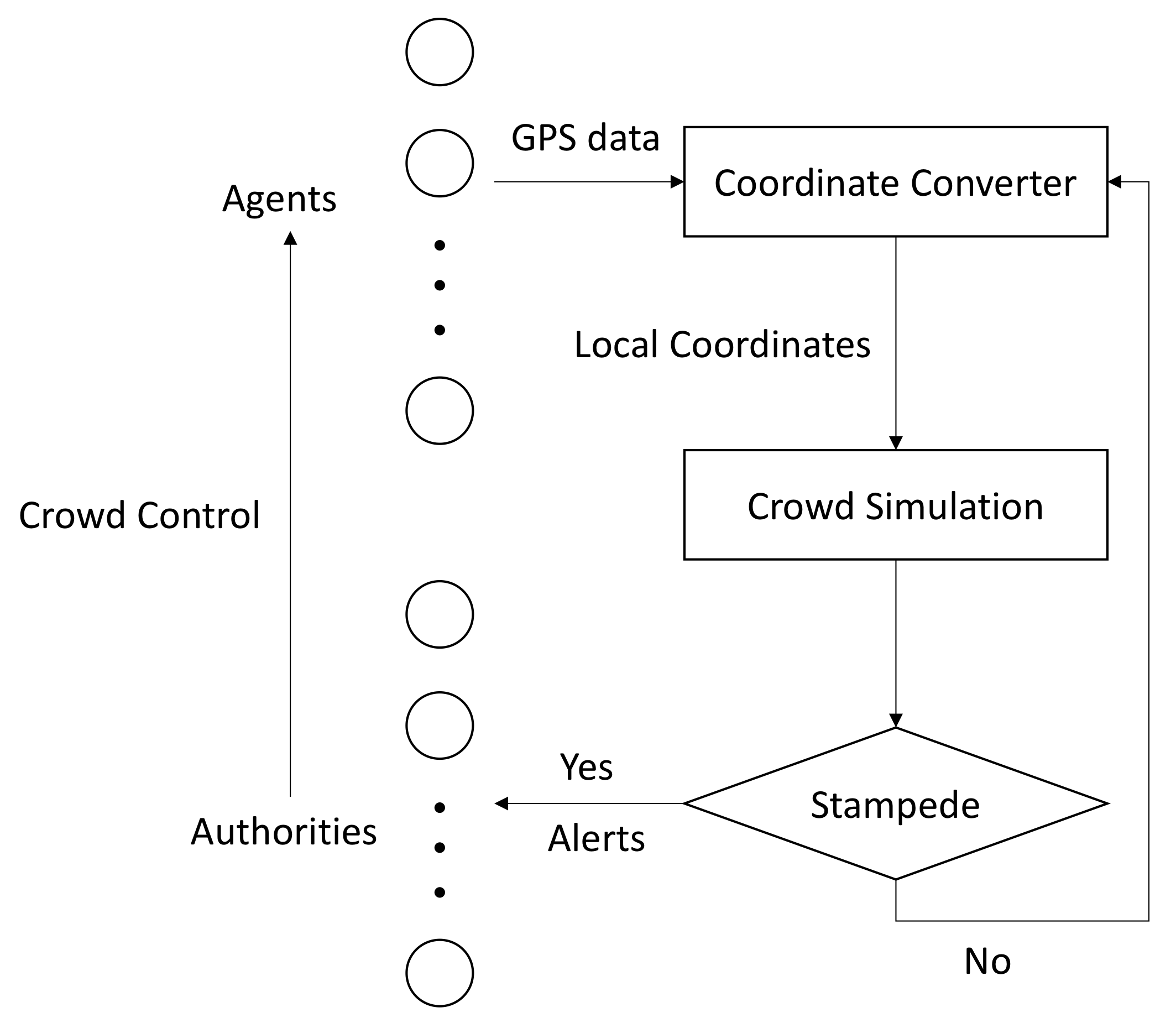}
	\caption{ARES steps to detect stampede accidents in large crowd events.}
	\label{fig:flowchart}
\end{figure}
Figure~\ref{fig:flowchart} outlines the major steps of ARES. The details of the steps are as follows.
\begin{enumerate}
    \item \textbf{Collect agents' current locations and velocities}. In the beginning of each interval $T$, we collect all agents' locations and velocities in the field. The locations and velocities are measured in the global coordinate system. 
    
    \item \textbf{Convert agents' locations and velocities to the local coordinate system}. Agents' longitude and latitude coordinates would be converted to local coordinates, which are relative locations in the venue map. In addition, agents' velocities in the global coordinate system would also be converted to the ones in the local coordinate system.  
    
    \item \textbf{Run the simulation with a pedestrian model and a stampede assessment method}. Agents' locations and velocities from Steps 2 are input parameters for the pedestrian model. We employ Monte Carlo method to run simulations and execute $M$ simulations in parallel using GNU parallel~\cite{tange2018gnu} (other parallelization techniques can be used as well). This is done to ensure that $K < T$ and the time difference is sufficient for on-site authorities to respond. 
    \item \textbf{Compute simulated stampede probabilities}. Each realization of simulation reports if a stampede has happened or not. After all the simulations complete, we calculate the expected value of the simulated stampede probability $p$ based on the Law of Large Numbers.
    
    \item \textbf{Send alerts to on-site authorities if necessary}. If the value of $p$ exceeds a threshold, an alert message can be sent out to notify on-site authorities, so that they could control the crowd near the specific spot.
    
    \item \textbf{Repeat the above steps once a new batch of locations and velocities data comes in.} Once a new batch of locations and velocities data has been collected after interval $T$, we repeat the procedure from Step 1 to Step 5.
\end{enumerate}

There are two important assumptions in ARES. First, we assume that each pedestrian has a wearable GPS device. This assumption is reasonable as mentioned in Section~\ref{sec:introduction}. Second, we assume that we know each agent's next destination based on the current state in the simulation. We believe that this is also a reasonable assumption in many cases. For instance, we know that pilgrims will stop and throw pebbles at three pillars on Jamarat bridge during Hajj, and soccer fans will move towards the nearest exit (or the one that they came in) to leave a stadium.

\subsection{Simulation framework}\label{subsec:simulation}%Menge

We employ Menge~\cite{curtis2016menge} as our crowd simulation framework. Menge has two built-in pedestrian models: ORCA and PedVO. The former is based on the Optimal Reciprocal Collision Avoidance (ORCA) algorithm~\cite{van2011reciprocal}, and the latter is based on the Pedestrian Velocity Obstacle (PedVO) model~\cite{curtis2014pedestrian}. As a successor to the ORCA model, PedVO preserves ORCA's geometric optimization technique and introduces a self-adaptive velocity technique by employing the fundamental diagram---the phenomenon that pedestrian speed reduces as density increases. The PedVO model captures both physiological and psychological factors that give rise to the fundamental diagram. It computes a collision-free velocity for each pedestrian based on a preferred velocity and its local neighborhood (i.e., other pedestrians and obstacles). 

In our work, we choose PedVO to simulate pedestrians' behavior, but others may choose the most suitable pedestrian model (e.g., ORCA or Johansson et al.'s model~\cite{johansson2007specification}) to address their cases. The default parameters of the PedVO model in Menge are as follows: agents are represented as filled circles with radius equals 0.19 m~\cite{littlefield2012metric}, agents' preferred velocity equals 1.04~m/s~\cite{callisaya2017cognitive}. Note that the default preferred velocity is lower than the average preferred walking speed---1.4~m/s~\cite{browning2006effects}. However, this setup is adequate for a crowd of people in a pilgrimage setting. The PedVO model uses the preferred velocity to compute an actual velocity based on the surroundings (the maximum velocity is 2~m/s~\cite{zkebala2012pedestrian}). The maximum acceleration rate is 5~m/s$^2$~\cite{alonso2014simulation}. Other calibration parameters of PedVO can be found in~\ref{sec:pedvo}. Note that these values may be adjusted if necessary (other reasonable values of the parameters are discussed in~\cite{dridi2015list}).

\begin{figure}[!ht]
	\centering
	\includegraphics[width=0.9\columnwidth]{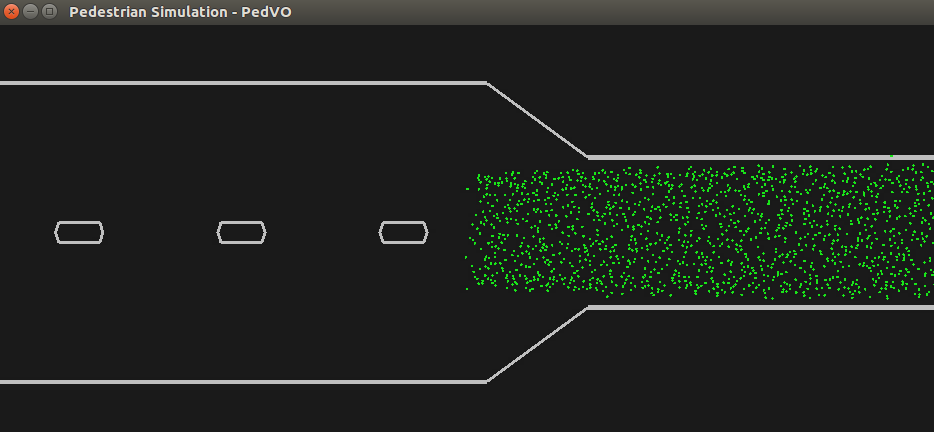}
	\caption{Simulation venue map. A set of emulated agents are generated on the right ramp (in a rectangular grid), and they are moving towards the left in the venue. The width of the ramps is 22 m, and the width of the bridge is 44 m. The three obstacles simulate the three pillars on the Jamarat bridge where pilgrims throw stones. Agents stop at each pillar for an average of 60 simulation time units (i.e., 60 s in our simulation), and they get off the bridge after that. Note that this venue is smaller than the actual Jamarat bridge, but it is sufficient to illustrate the concept.}
	\label{fig:menge}
\end{figure}
To simulate a representative crowd congestion scenario, we build a simplified venue map of the Jamarat bridge based on the map in the supplementary material of Helbing et al.'s work~\cite{helbing2007supplementary}. The Jamarat bridge is the place where the stampede happened during Hajj in 2006. A simulated scene on the map can be seen in Fig.~\ref{fig:menge}. 

\subsection{Stampede assessment methods}\label{subsec:stampede}

We extend Menge with three stampede assessment methods. The three threshold-based stampede assessment methods that we employ are crowd pressure method~\cite{helbing2007dynamics}, physical force method~\cite{elliott1993football, helbing2002simulation}, and density method~\cite{lee2005exploring}. Details of these methods are given below.

\subsubsection{Crowd pressure method}  
As crowd density reaches a certain dangerous level (typically seven or eight people per square metre~\cite{helbing2007dynamics}), the physical contacts transfer and accumulate among the crowd, and individual motion gets replaced by mass motion. When the pressure becomes too large, the mass splits up into different clusters, which causes strong variations of local strengths and velocities. This can lead to sudden and uncontrolled pressure release. Such a phenomenon is referred to as crowd turbulence. Crowd turbulence is observed by analyzing video recordings of the crowd disaster during the Hajj in 2006~\cite{helbing2007dynamics}. People try to escape such a situation; therefore, they push each other. This is called crowd panic as per~\cite{helbing2007dynamics}\footnote{Note that a stampede may also occur in the absence of panic.}. The crowd panic introduces extra power into the compressed crowd. Eventually, stampede accidents occur when a significant amount of power accumulates and then releases.

Both crowd turbulence and stampede accident can be quantified by crowd pressure (which is different from physical pressure). Crowd pressure is defined as local density times the local velocity variance~\cite{helbing2007dynamics}. 

Formally, the local density at a given place $\vec{r}=(x,y)$ and time $t$ is defined as
\begin{equation}\label{eq:density}
\rho (\vec{r}, t) = \sum_j f\Big(\vec{r}_j (t) - \vec{r}\Big),
\end{equation}
where $\vec{r}_j(t)$ is the position of pedestrian $j$ in the surrounding of $\vec{r}$ at time $t$, and $f\Big(\vec{r}_j (t) - \vec{r}\Big)$ is a Gaussian distance-dependent weight function
\begin{equation}\label{eq:weight}
f\Big(\vec{r}_j (t) - \vec{r}\Big) = \frac{1}{\pi R^2} \exp \left[- \| \vec{r}_j (t) - \vec{r} \|^2 / R^2\right],
\end{equation}
where $R$ defines the radius of the local measured area. The local velocity at position $\vec{r}$ and time $t$ is defined as
\begin{equation}\label{eq:speed}
\vec{V} (\vec{r}, t) = \frac{\sum_j \vec{v}_j f\Big(\vec{r}_j (t) - \vec{r}\Big)}{\sum_j f\Big(\vec{r}_j(t) - \vec{r}\Big)},
\end{equation}
where $\vec{v}_j$ is the velocity of of pedestrian $j$ in the surrounding of $\vec{r}$ at time $t$. 
Then, we can determine the local crowd pressure based on Eqs.~(\ref{eq:density}) and (\ref{eq:speed}) as
\begin{equation}\label{eq:pressure}
P(\vec{r}, t) = \rho (\vec{r}, t) \cdot \Var_{\vec{r},t} (\vec{V}),
\end{equation}
where $\Var_{\vec{r},t}(\vec{V})$ is the local velocity variance at time $t$, and its value is calculated by
$\Var_{\vec{r}, t} (\vec{V}) = \langle [ V(\vec{r}, t) - \langle V \rangle_{\vec{r}} ]^2 \rangle_{\vec{r}}$,
where $\langle V \rangle_{\vec{r}}$ is the average local velocity over the circular area centered in $\vec{r}$ and with radius $R$. According to~\cite{helbing2007dynamics}, crowd turbulence starts when the pressure reaches $0.02$ s$^{-2}$, and stampede starts when the pressure reaches $0.04$ s$^{-2}$.

% force model
\subsubsection{Physical force method} 
The second stampede assessment method is based on physical force~\cite{elliott1993football, helbing2002simulation}. The idea is that physical interactions in jammed crowds add up and can lead to  dangerous consequences. In simulations, we monitor the physical force among agents, and set a threshold for the force. A stampede is measured if any monitored physical force exceeds the threshold. Pure force function is defined by Newton's second law, $F = m \cdot a$, where $m$ is the mass, and $a$ is the acceleration. Based on the research in~\cite{elliott1993football, helbing2002simulation}, 4500 N is the force that, on average, may cause an adult to fall on the ground, potentially causing stampede accidents. 

For parameterization, we set $m$ to follow a normal distribution with mean value of 70 kg (which is a typical weight of adults in Europe~\cite{walpole2012weight}) and standard deviation of 10 kg~\cite{launer1996weight}. We set the minimum value of $m$ to 50 kg, and the maximum value of $m$ to 100 kg. Note that these values can be adjusted according to different nationalities and occasions. The value of $a$ can be calculated by
\begin{equation}\label{eq:acceleration}
a = \frac{\vec{v}(t + \Delta t) - \vec{v}(t)}{\Delta t},
\end{equation}
where $\Delta t$ is the change in time, and $\vec{v}(t)$ is the velocity at time $t$.

\subsubsection{Density method} 
Our last stampede assessment method is based on a local density threshold. Lee and Hughes~\cite{lee2005exploring} found the correlation between crowds involving stampede accidents and crowd density. Typical stampede accidents involve a large number of pedestrians, which is generally about seven or eight pedestrians per square metre (possibly up to 13 pedestrians per square metre). The maximum observable densities are naturally dependent on different nationalities~\cite{lee2005exploring}. 

\begin{figure}[!ht]
	\centering
	\includegraphics[width=0.9\columnwidth]{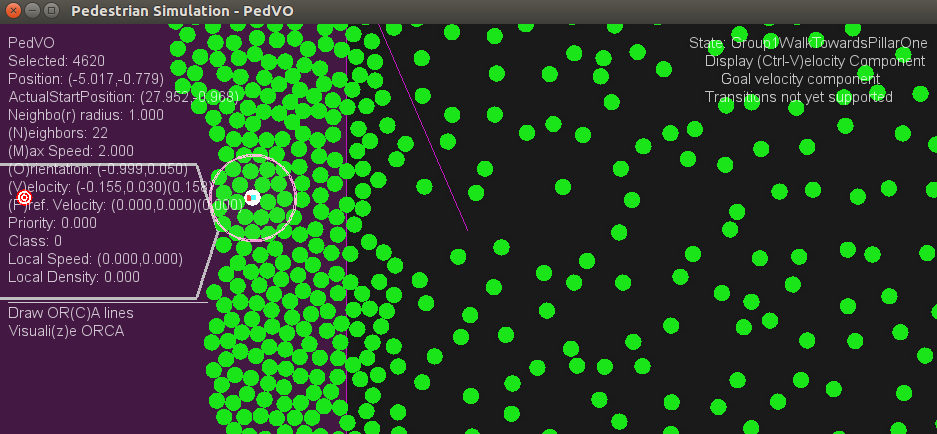}
	\caption{A stampede is observed in the density method. The white filled circle is the agent whose neighbors exceed the threshold---22, and the neighbors are within the unfilled circle with $R = 1$ m. All the information about the agent and its neighborhood can be seen in the left part of the figure.}
	\label{fig:density-sample}
\end{figure}
Compared to the local density calculation (which is an estimate) in the crowd pressure method, we employ a classical density estimation. We monitor the number of neighbors (local density) for each agent during simulations, and set a threshold for the local density, which is calculated as $D = L/(\pi R^2)$, where $L$ is the number of neighbors within the area where the radius is $R$. In our experiments, we set the density threshold to seven agents per square metre as per~\cite{lee2005exploring}. A stampede is measured if any local density exceeds this threshold. 

Figure~\ref{fig:density-sample} shows a scenario where a stampede is detected with the density threshold. In this case, we have $L = 22$ agents and $R = 1$ m. Thus, $D \approx 7.01$ m$^{-2}$. In our experiments, the maximum density that we observed is 29 agents within the radius of one metre, i.e., $D \approx 9.24$ m$^{-2}$. 

In this method, the density calculation does not take into account the obstacles (i.e., pillars and walls). We conjecture that if we took obstacles into account (when computing local densities), the simulated stampede probabilities might increase, making the estimates more sensitive (i.e., increasing false positive error rate). As the coordinates of agents and obstacles are known, one may create an algorithm to estimate the impact of obstacles on the local densities.  

In the literature, there exist various approaches to estimate local or global densities~\cite{marana1999estimating, steffen2010methods, tordeux2015quantitative}, but they are not applicable to our experiments. A fractal-based image processing technique was proposed to estimate crowd density in~\cite{marana1999estimating}. The work of \cite{steffen2010methods} focused on estimating global density rather than local density. The method in~\cite{tordeux2015quantitative} can be used to estimate local density, but it is not applicable to our setup, because it assumes that local densities are homogeneous (which is violated in our case). These methods cannot be directly applied to our case, but they can be extended to take into consideration only the accessible areas. For example, we know all the coordinates of obstacles and agents in our case study, and we can remove inaccessible areas from the density estimators. Because the standard approach is sufficient for the density estimation, we leave the research of obstacle-aware density estimation to future work.

\subsection{Noise modeling}\label{subsec:noise}
Now, we introduce the noise modeling technique in crowd simulation. In real world, personal GPS devices often report locations with certain amount of errors. These errors include horizontal distance errors, vertical distance errors, and velocity errors. In this paper, we only consider horizontal errors as our map is two-dimensional. Velocity errors may have impact on the two velocity-dependent measurements, i.e., the crowd pressure method and the physical force method; however, these errors will not affect the density method (which is velocity independent). More importantly, it was shown that velocity errors are negligible\footnote{The velocity estimate can achieve the accuracy of centimeters per second by using Doppler frequency shifts of the received signal produced by user-satellite motion~\cite{petovello2015does}. In GPS applications, improved performance is achieved by processing differences of consecutive carrier phase measurements, which are better than the ``raw'' Doppler measurements~\cite{petovello2015does}.} compared to distance errors: velocity errors are in the order of centimetres while distance errors are in the order of metres~\cite{wang2008error}. Therefore, we chose to ignore the velocity error.  

The empirical distribution of latitude and longitude errors is very similar to normal distribution~\cite{zandbergen2008positional,zelenkov2008accuracy}. Consequently, the GPS horizontal error, which is geometric sum of these two errors, is very similar to Rayleigh distribution~\cite{papoulis2002probability}. The Cumulative Distribution Function (CDF) for Rayleigh distribution is:
\begin{equation}\label{eq:cdf2}
P(Z \leq z) = 1 - \exp\left(-\frac{z^2}{E^2}\right), \quad E > 0,
\end{equation}
where $z$ is the distance error, and $E$ is the root-mean-squared distance error. 

Finally, we could use Eq.~(\ref{eq:cdf2}) to generate random errors $z$ to add to the starting true positions of agents in the simulation. This simulates the scenario where the real-time simulation starts with GPS position errors. In our simulations, we vary the values of $E$ from zero to ten metres to test the vulnerability of the stampede assessment method. As mentioned earlier, Zandbergen and Barbeau's research shows that the positional error of smartphones ranges from five to eight metres~\cite{zandbergen2011positional}. Wilson's tests on personal GPS devices suggest a similar estimate~\cite{wilson2002gps}. Thus, we believe that it is reasonable to set the range of the values of $E$ as zero to ten metres, so that we could cover most of the cases. Note that in the case when $E=0$, we use the actual starting positions of the agents; thus, we do not need to use Eq.~(\ref{eq:cdf2}) in this case.

% noise figures
\begin{figure}[!ht]
	\centering
	\includegraphics[width=0.9\columnwidth]{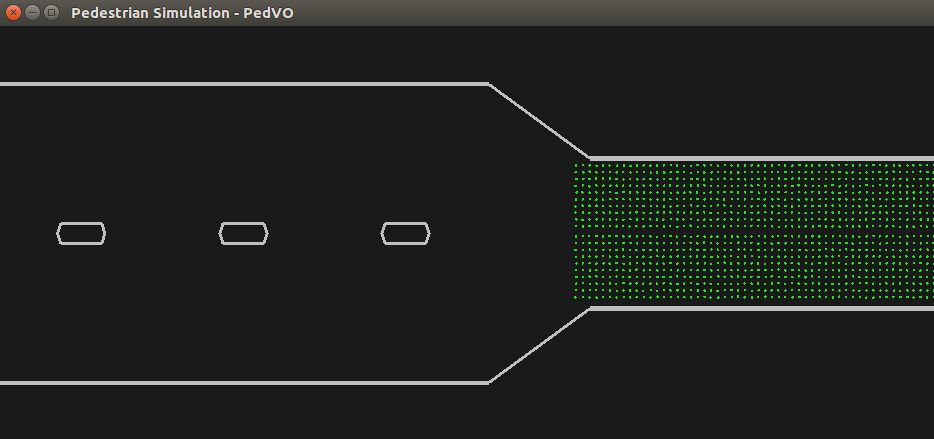}
	\caption{Starting positions with $E = 0$ m.}
	\label{fig:noise0}
\end{figure}
\begin{figure}[!ht]
	\centering
	\includegraphics[width=0.9\columnwidth]{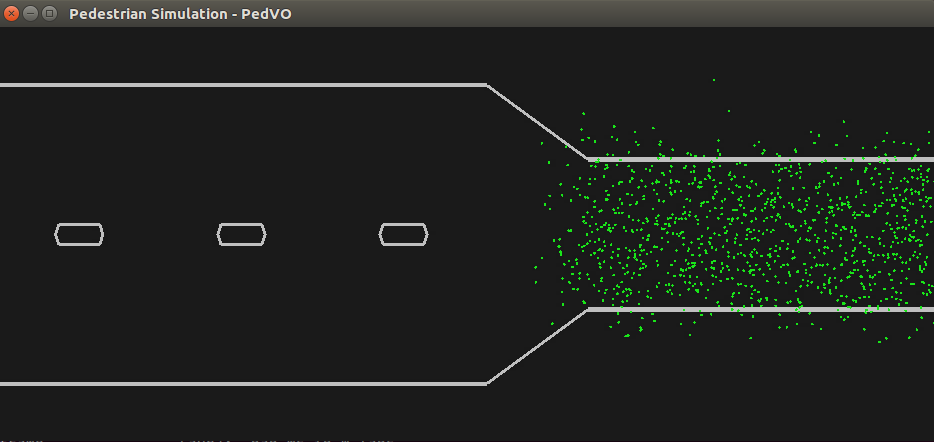}
	\caption{Starting positions with $E = 5$ m.}
	\label{fig:noise5}
\end{figure}
\begin{figure}[!ht]
	\centering
	\includegraphics[width=0.9\columnwidth]{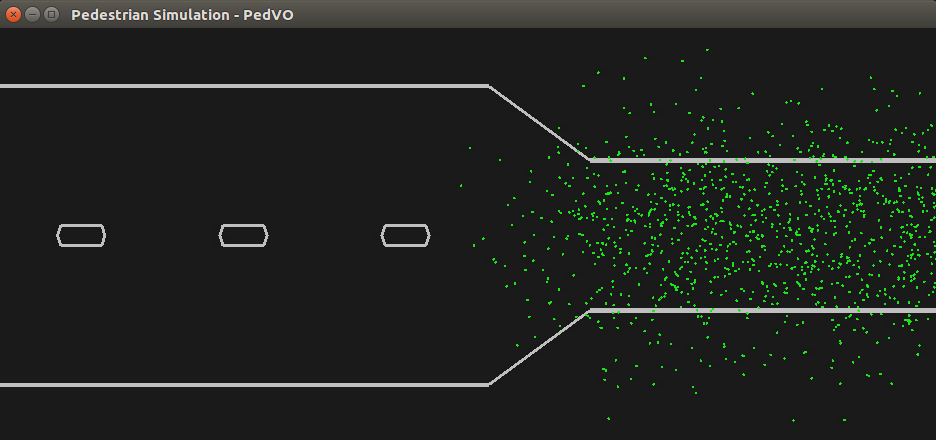}
	\caption{Starting positions with $E = 10$ m.}
	\label{fig:noise10}
\end{figure}

Figure~\ref{fig:noise0} shows an example, where agents' starting positions have no noise added ($E = 0$). Figures~\ref{fig:noise5} and~\ref{fig:noise10} show two examples, where noise is added to starting positions. The values of noise are drawn from the Rayleigh distribution as per Eq.~(\ref{eq:cdf2}) with $E = 5$ and $10$ m, respectively. As can be seen in Figs.~\ref{fig:noise5} and~\ref{fig:noise10}, some agents are shifted outside of the bridge due to the errors. We deliberately choose not to alter agents' behavior in such simulations to assess the effect of the spread in simulated stampede probabilities (further discussion can be found in Section~\ref{subsec:pressure}). 

\subsection{Simulated stampede probabilities}\label{subsec:probability}
To estimate the simulated stampede probabilities and the impact of position errors, we employ Monte Carlo method. Given probabilistic nature of Menge simulation, every realization of the trajectories of agents from start to finish will be different. By the Law of Large Numbers, once we have sufficient number of realizations of the trajectories, we will be able to compute the expected value of the probability of simulated stampede along with the confidence interval for the expected value.

Every realization of the simulation will report if a stampede has occurred in a given run or not. Let us denote the outcome of the simulation by the indicator variable $I$, and the total number of realizations is $n$. If a stampede occurred in a given realization, $I$ will be equal to one, otherwise---to zero. Then, the expected value of the probability of simulated stampede $p$ is given by
\begin{equation}\label{eq:probability}
p = \frac{\sum_{j=1}^{n} I_j}{n}.
\end{equation}
The $95\%$ confidence interval of this expected value is 
\begin{equation}\label{eq:ci}
\approx [p+1.96\sigma/\sqrt{n},p-1.96\sigma/\sqrt{n}],
\end{equation}
where $\sigma$ is the standard deviation of the realizations of $I$. 

\subsection{Discussion}\label{subsec:discussion1}

As discussed in Section~\ref{sec:related}, there exists a large number of stampede assessment methods. For this study, we chose three of these methods. The design of this study is based on the concept of the critical case~\cite{yin2013case}. That is, the selected models are representative instances of the stampede assessment methods. While we cannot generalize these results to other methods,  the fact that these models are all affected by the measurement error suggests that other models may be affected as well. The same experimental examination can also be applied to other stamped prediction models and methods with well-designed and controlled experiments.

The GPS accuracy can be improved using various techniques, e.g., A-GPS, accelerometers, the Kalman filter, and the moving average filter (as discussed in Section~\ref{sec:related}). However, the resulting accuracy of the measurements will not be perfect. To illustrate this, we evaluate the performance of the standard Kalman filter applied to the noisy coordinates of the agents. We chose this particular filter, as it has been used for decades to eliminate GPS position noise~\cite{brown1992introduction}. 

Details of the analysis are given in \ref{sec:canceling}. In short, we observe that the standard Kalman filter reduces the noise significantly when an agent is following a monotonic trajectory (which has been observed in the past~\cite{moussakhani2014change}). However, the filter has difficulty in detecting an abrupt change in the direction or velocity of the move, which replaces random error with systemic error~\cite{gustafsson2000adaptive}. We also show that, when applied to a set of agents, the average position error (after application of the Kalman filter) ranges from 2.6 to 3.0 m. Moreover, the Kalman filter amplifies noise when the position errors are small, making it impractical for such a case. To conclude, the error after the adoption of the Kalman filter is still significant and may affect model performance. Therefore, let us explore how the noise affect the performance of the models.

\section{Experimental results}\label{sec:experiments}

We set up a testbed on our server, which has Ubuntu (4.4.0-103-generic) system, 56 CPUs, and 512 GB memory. We run simulations in the scene given in Fig.~\ref{fig:menge}; we use GNU parallel~\cite{tange2018gnu} to parallelize the simulations. We vary the values of the root-mean-squared distance error ($E$) and the number of agents ($N$), choose a sufficient number for repeated realizations (in our case, $n = 1000$), and calculate the expected value of probabilities ($p$), as well as the $95\%$ confidence interval ($C$) as follows.
\begin{enumerate}
	\item For $E = 0, 1, 2, \ldots, 10$ m and $N = 1, 5, 10, \ldots, 10240$ ($N$ increases by a factor of two, starting from $N=5$), repeat 1000  simulations with the same set of $\{E, N\}$. For each simulation, return one if stampede occurs; otherwise, return zero.
	\item Upon the completion of simulations with one set of $\{E, N\}$, compute $p$ and $C$ using Eqs.~(\ref{eq:probability}) and (\ref{eq:ci}), respectively.
\end{enumerate}

The pseudocode of this process is shown in Algorithm~\ref{alg:script}.

\begin{algorithm}[!ht] % [H] avoids the environment to float
	\caption{Pseudocode to execute the simulation}\label{alg:script}
	\begin{algorithmic}[1] %[1] put number everyline
		\Require $E, N$
		\Ensure $p, C$
		\For{each stampede assessment method} \Comment iterate over $R$ if needed
		\For{$E = 0, 1, \ldots, 10$}
		\For{$N = 1, 5, 10, \ldots, 10240$}
		\State $i\gets 1$
		\Repeat
		\State Run simulation for a given value of $E$ and $N$
		\If{stampede occurs}
		\State Return 1
		\Else
		\State Return 0
		\EndIf
		\State $i\gets i+1$
		\Until{$i>1000$}
		\State Compute $p$ and $C$ using Eqs.~(\ref{eq:probability}) and (\ref{eq:ci})
		\EndFor
		\EndFor
		\EndFor
\end{algorithmic}
\end{algorithm}

\subsection{Crowd pressure method}\label{subsec:pressure}

\begin{figure}[!ht]
    \centering
	\begin{subfigure}[b]{0.49\columnwidth}
	\centering
		\includegraphics[width=1.0\columnwidth]{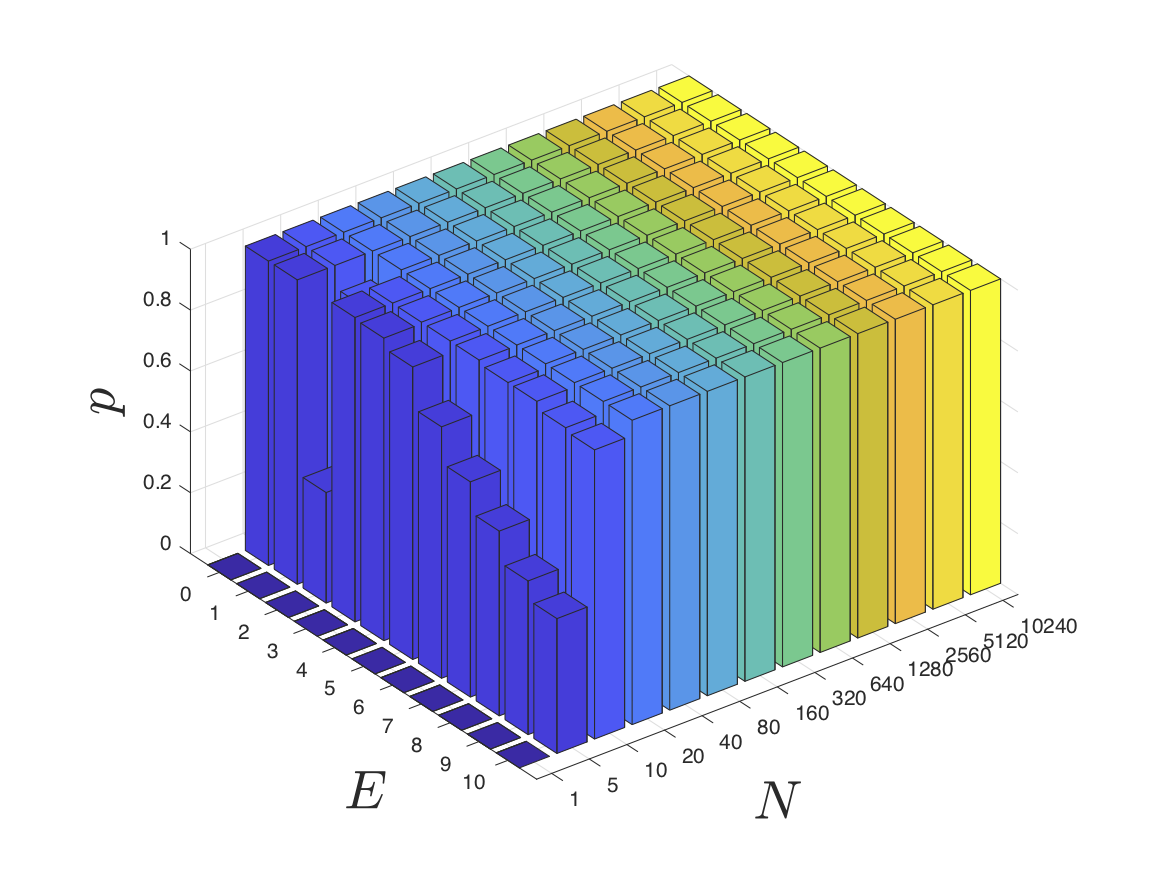}
		\caption{$R = 1$ m}
		\label{fig:pressure-prob-1}
	\end{subfigure}
	\begin{subfigure}[b]{0.49\columnwidth}
	\centering
		\includegraphics[width=1.0\columnwidth]{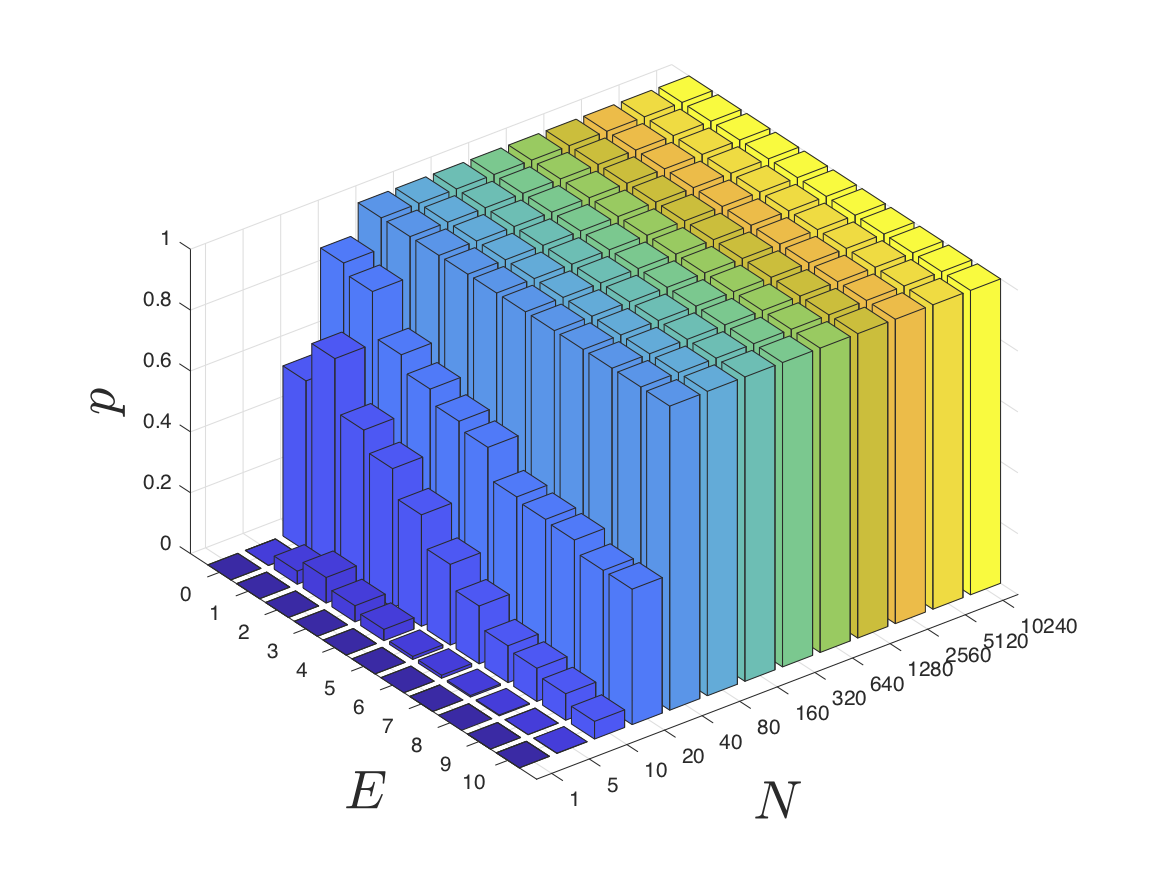}
		\caption{$R = 2$ m}
		\label{fig:pressure-prob-2}
	\end{subfigure}
	
	\begin{subfigure}[b]{0.49\columnwidth}
	\centering
		\includegraphics[width=1.0\columnwidth]{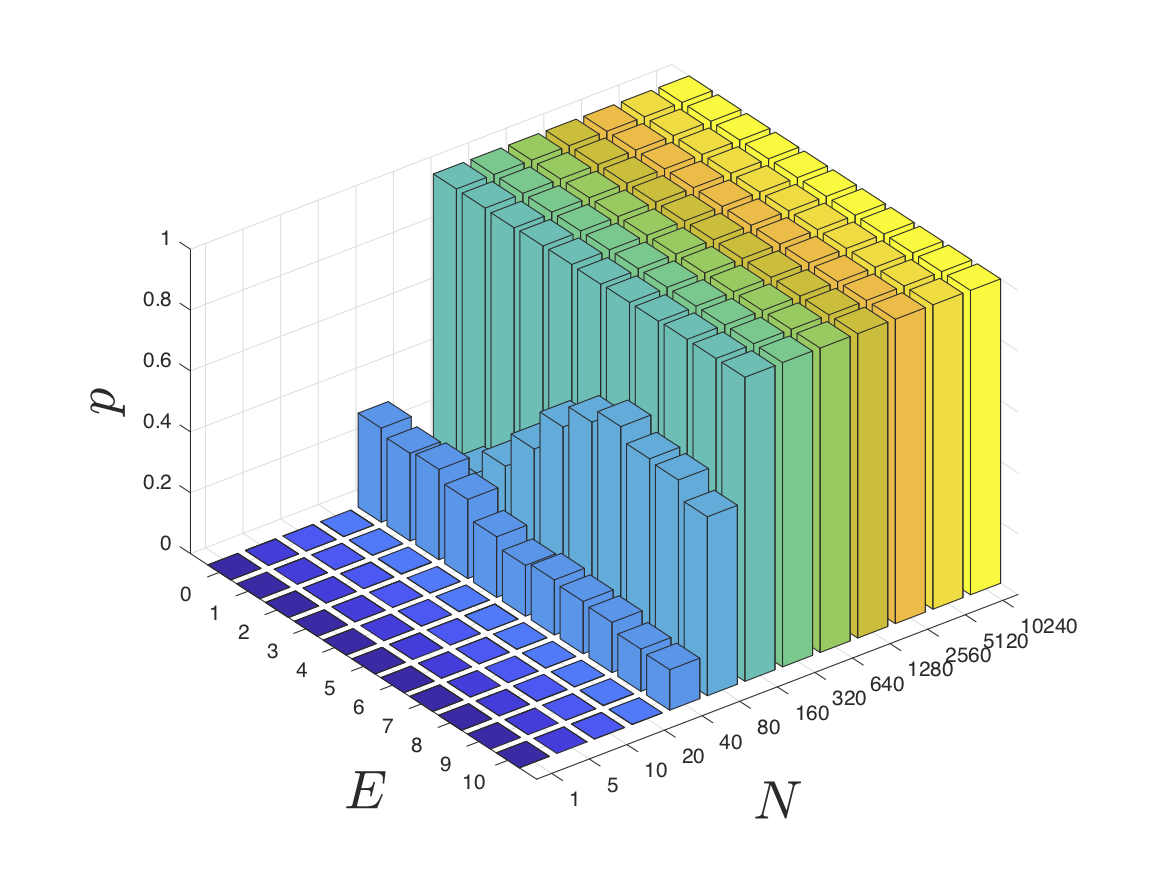}
		\caption{$R = 3$ m}
		\label{fig:pressure-prob-3}
	\end{subfigure}
	\begin{subfigure}[b]{0.49\columnwidth}
	\centering
		\includegraphics[width=1.0\columnwidth]{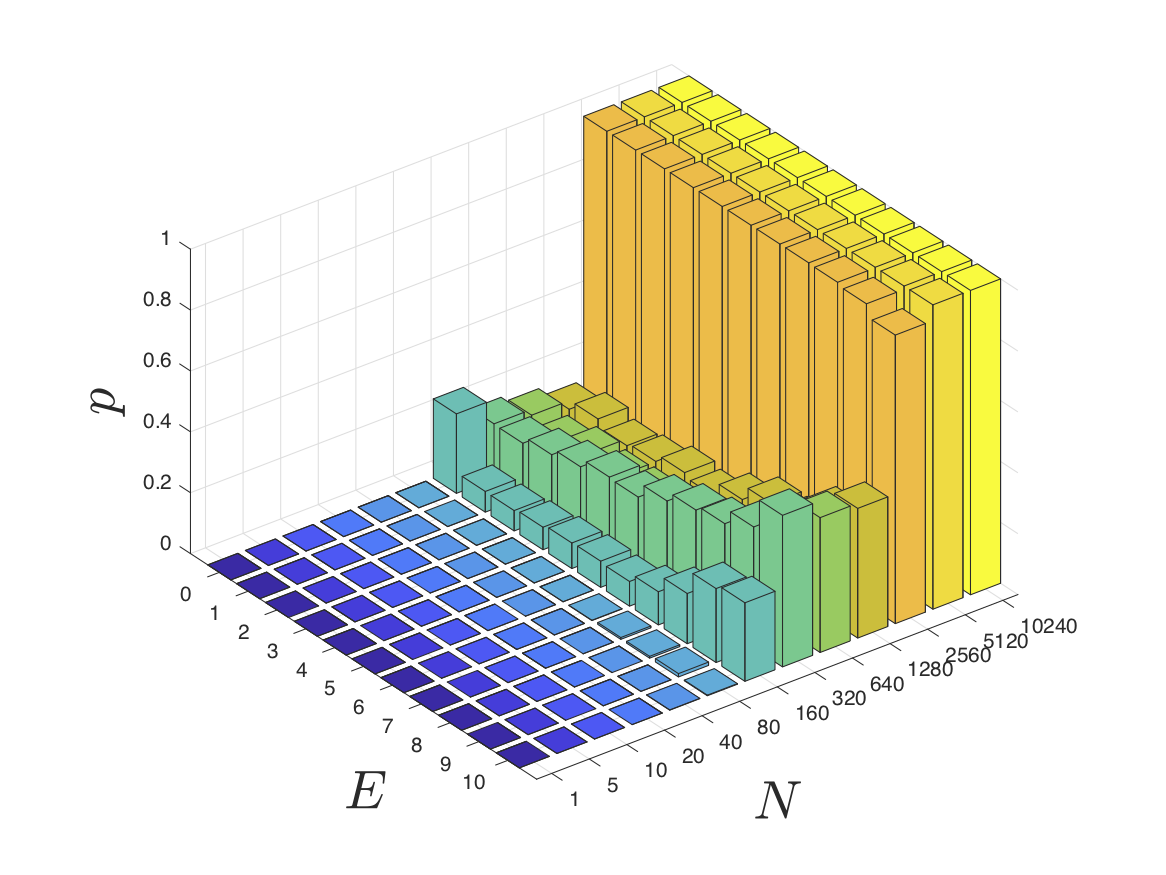}
		\caption{$R = 4$ m}
		\label{fig:pressure-prob-4}
	\end{subfigure}
	\caption{Stampede probabilities in the pressure method. Three-dimensional bar graph of $(N,E,p)$-tuples. Stampede probabilities with (a) $R = 1$ m, (b) $R = 2$ m, (c) $R = 3$ m, and (d) $R = 4$ m.}
	\label{fig:pressure-prob}
\end{figure}

Experimental results of simulated stampede probabilities from the crowd pressure method are shown in Fig.~\ref{fig:pressure-prob}. In this method, the values of the crowd pressure depend on the choice of $R$ in Eq.~(\ref{eq:weight}). Figures~\ref{fig:pressure-prob-1},~\ref{fig:pressure-prob-2},~\ref{fig:pressure-prob-3}, and~\ref{fig:pressure-prob-4} represent the simulated stampede probabilities when $R = 1$, $2$, $3$, and $4$ m, respectively. The threshold of the pressure is set to $0.04$ s$^{-2}$; if we detect a value of the pressure above the threshold---we record a stampede.  In the experiments, we test the sensitivity of this metric to the changes of $R$ and $N$. From Fig.~\ref{fig:pressure-prob}, we have four observations as follows.
\begin{itemize}
	\item The crowd pressure method is very sensitive when $R = 1$ m. It generates false positive errors when the number of agents is relatively small: even with 20 agents, the crowd pressure method provides stampede prediction with $p=1$ (almost surely). %Such a stampede method could trigger false positive errors.
	\item The crowd pressure method becomes less sensitive when $R$ increases: the greater the $R$---the closer the local density $\rho$ to the average of the global density. As $R$ grows large enough to cover the whole area which includes all the agents, all local densities will result in the same value, which corresponds to the overall number of agents divided by the area.
	\item When the values of $R$ and $E$ are fixed, the value of $p$ tends to increase with the growth of $N$. However, this is not always the case (as shown in Figs.~\ref{fig:pressure-prob-3} and~\ref{fig:pressure-prob-4}). The underlying reason is that the pedestrian model---PedVO---uses the fundamental diagram to model the correlation between velocities and surroundings. Less neighbors means higher velocities in the beginning of simulations. Consequently, agents may have to change velocities dramatically when facing the pillars. As a result, the variance of local velocity in Eq.~(\ref{eq:pressure}) increases. Experiments with different distributions of agents' starting positions have been done (but not presented here to avoid repetition), where the $p$ value can increase with the value of $N$ monotonically.
	\item Position errors may affect the $p$ values, especially when $0<p<1$. The impact of position errors could be positive or negative. For example, in Fig.~\ref{fig:pressure-prob-3}, increased position errors decrease the simulated stampede probability when $N=40$. However, the $p$ value rises and then drops with the growth of $E$ when $N=80$. The reason that the $p$ value decreases, when $E$ increases from seven to ten, is that the crowd is spread out when the position errors are introduced (as shown in Figs.~\ref{fig:noise5} and~\ref{fig:noise10}). Thus, physical contacts among agents are generally small when $E$ is large. However, the spreading-out positions may randomly generate some ``hot zones'', and this explains why the $p$ value increases in some cases mentioned above.
\end{itemize}

\begin{figure}[!ht]%[H]
	\centering
	\begin{subfigure}[b]{0.49\columnwidth}
	\centering
		\includegraphics[width=\columnwidth]{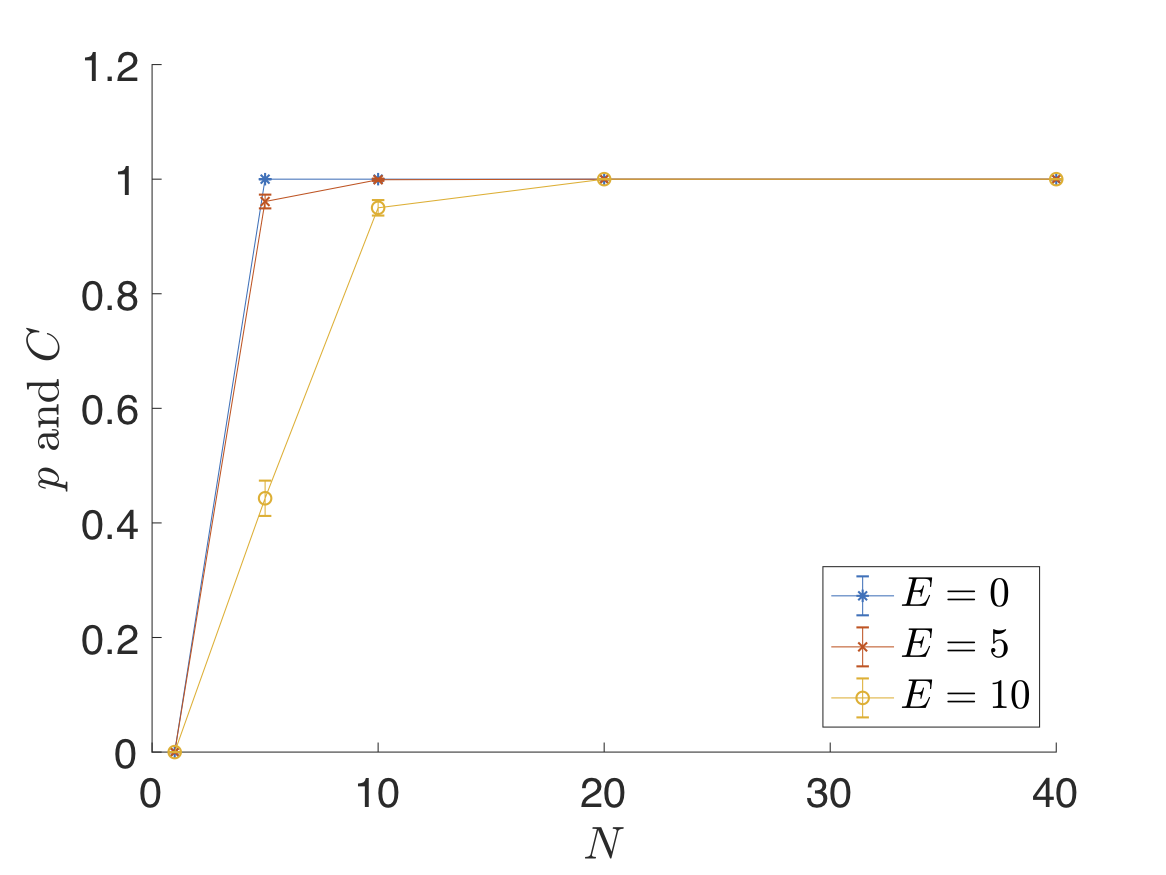}
		\caption{$R = 1$ m}
		\label{fig:r1-ci}
	\end{subfigure}
	\begin{subfigure}[b]{0.49\columnwidth}
	\centering
		\includegraphics[width=\columnwidth]{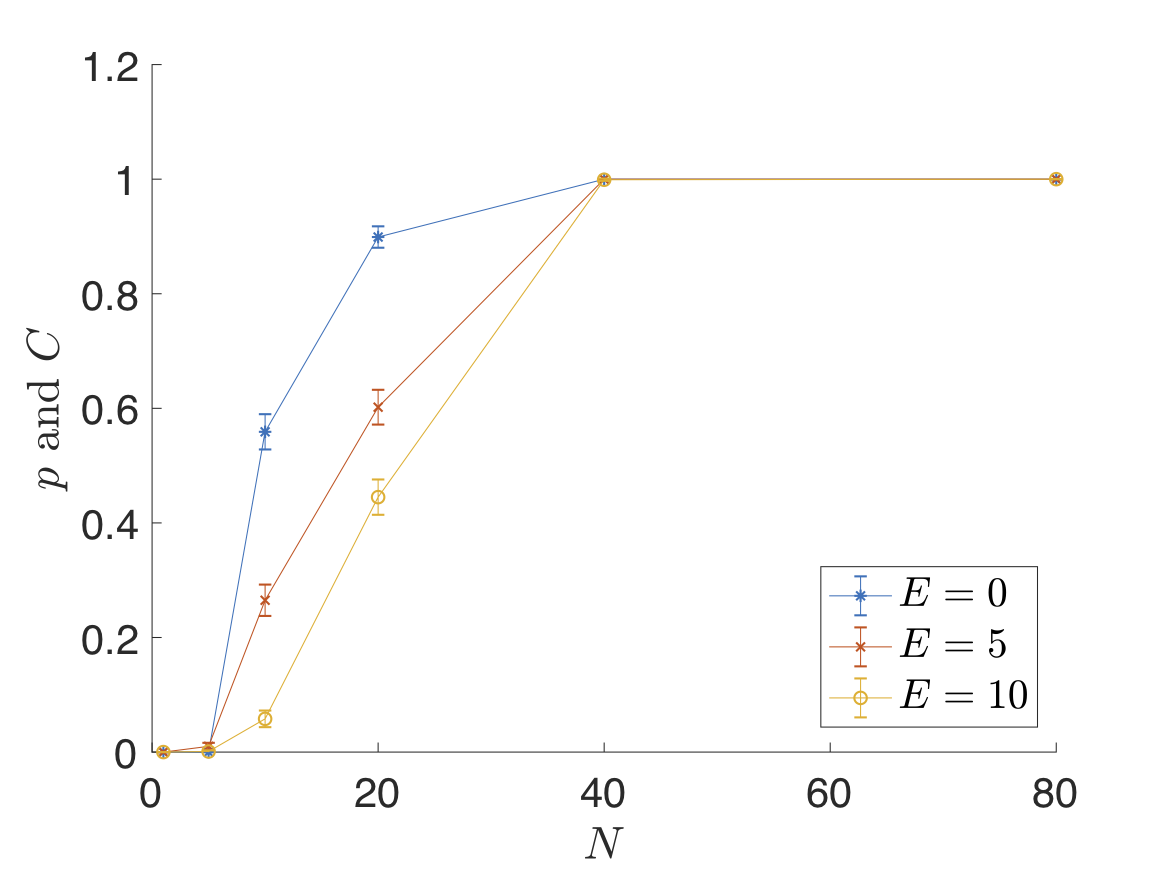}
		\caption{$R = 2$ m}
		\label{fig:r2-ci}
	\end{subfigure}

	\begin{subfigure}[b]{0.49\columnwidth}
	\centering
		\includegraphics[width=\columnwidth]{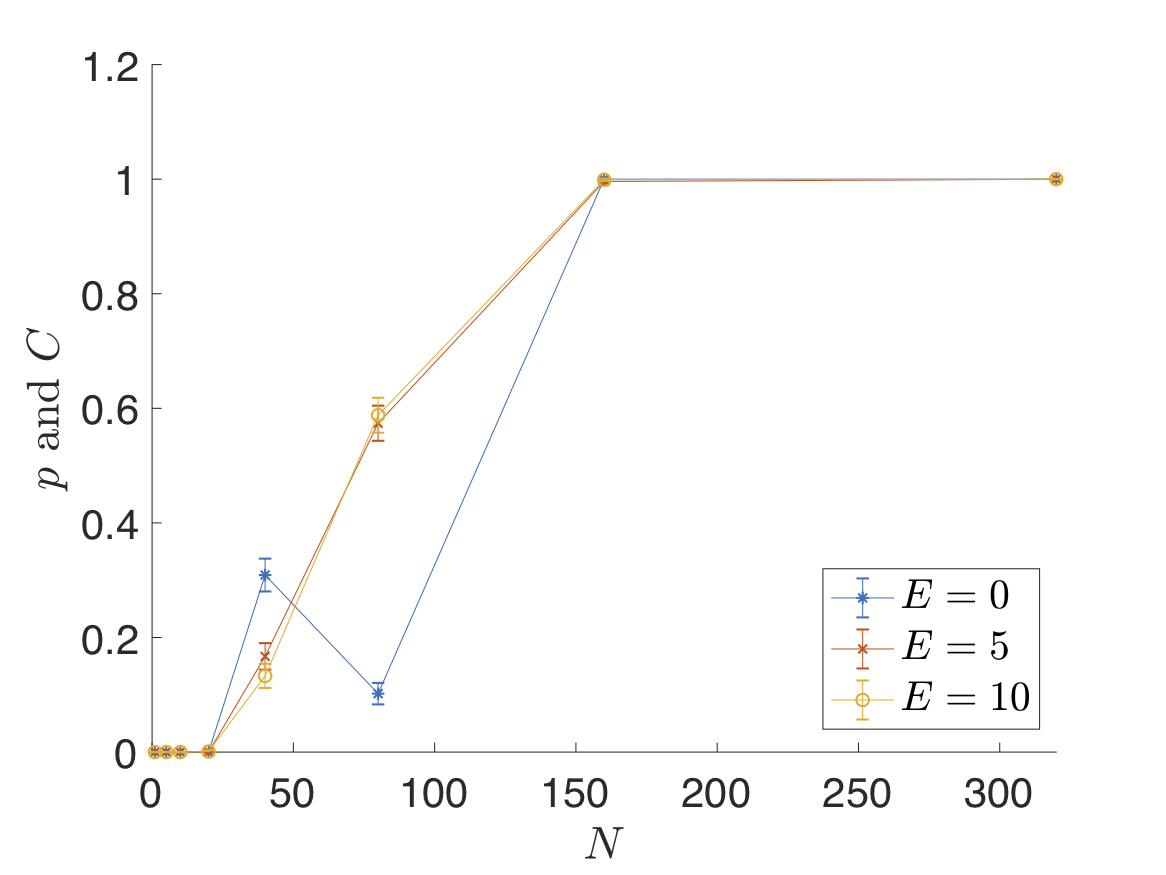}
		\caption{$R = 3$ m}
		\label{fig:r3-ci}
	\end{subfigure}
	\begin{subfigure}[b]{0.49\columnwidth}
	\centering
		\includegraphics[width=\columnwidth]{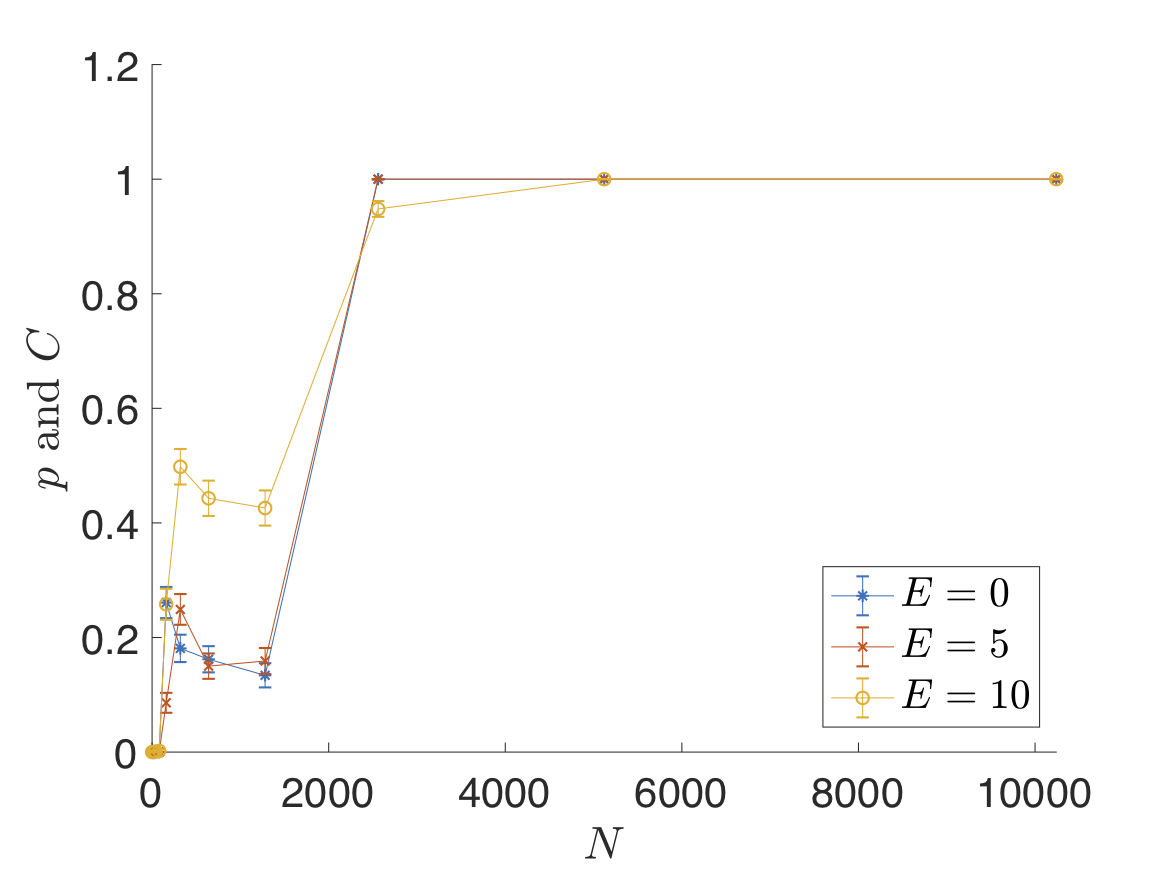}
		\caption{$R = 4$ m}
		\label{fig:r4-ci}
	\end{subfigure}
	\caption{Simulated stampede probabilities with confidence levels in the crowd pressure method. Two-dimensional line plots of simulated stampede probabilities ($p$) and confidence levels ($C$). Values of $p$ are presented in dots connected by lines, while values of $C$ are presented in bars around the corresponding values of $p$. Plots of $E=0$ m are in blue, plots of $E=5$ m are in red, and plots of $E=10$ m are in yellow. The values of $p$ and $C$ with (a) $R=1$ m and $N=1,\ldots,40$, (b) $R=2$ m and $N=1,\ldots,80$, (c) $R=3$ m and $N=1,\ldots,320$, (d) $R=4$ m and $N=1,\ldots,10240$.}
	\label{fig:pressure-ci}
\end{figure}

To estimate the confidence intervals of simulated stampede probabilities in the crowd pressure method, we show the confidence intervals calculated by Eq.~(\ref{eq:ci}) with $R = 1$, 2, 3, and 4 m in Fig.~\ref{fig:pressure-ci}. To improve comprehension of the plots, we choose three representative values of $E$; namely, $E=0$, $5$, and $10$ m. Besides the observations from Fig.~\ref{fig:pressure-prob}, we have two new observations from Fig.~\ref{fig:pressure-ci} as follows.
\begin{itemize}
	\item The confidence interval is relatively narrow, which means that the sampling size of the method is reasonable and the estimated stampede probabilities are precise. In other words, we are confident that we would get a similar result when re-running the simulations. 
	\item Generally speaking, the larger the position errors we observe, the less accurate the estimated stampede probabilities are. For example, the two curves of $E = 0$ m and $E = 5$ m are close to each other in Fig.~\ref{fig:r1-ci}, while the curve of $E=10$ m departs from them.
\end{itemize}

\subsection{Physical force method}\label{subsec:physical}

\begin{figure}[!ht]
	\centering
	\begin{subfigure}[b]{0.49\columnwidth}
		\includegraphics[width=\columnwidth]{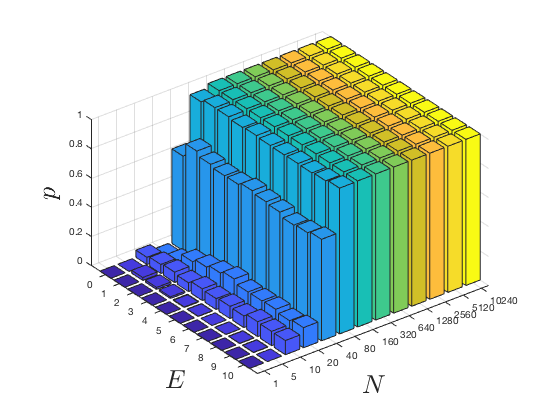}
		\caption{Force method}
		\label{fig:force-prob}
	\end{subfigure}
	\begin{subfigure}[b]{0.49\columnwidth}
		\includegraphics[width=\columnwidth]{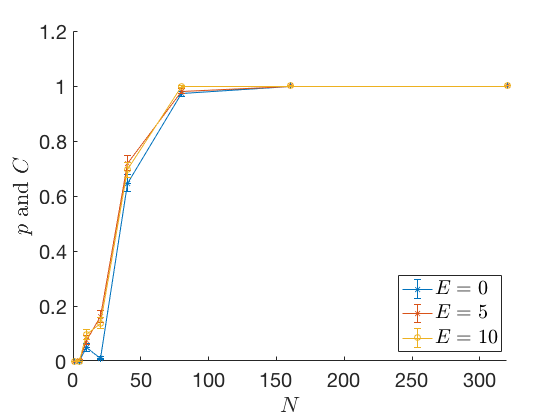}
		\caption{Force method}
		\label{fig:force-ci}
	\end{subfigure}
	\caption{Simulated stampede probabilities and the corresponding confidence levels in the physical force method. (a) Simulated stampede probabilities of the physical force method in three-dimensional bar graph of $(N,E,p)$-tuples. (b) Values of $p$ and $C$ of the physical force method in two-dimensional line plot.}
	\label{fig:force}
\end{figure}

Let us now present the experimental results for the physical force method. Note that unlike the previous method, it does not need to set the $R$. In this method, the threshold is 4500 N. The simulated stampede probabilities can be seen in Fig.~\ref{fig:force-prob}. We also plot the simulated stampede probabilities and confidence intervals for the physical force method in Fig.~\ref{fig:force-ci}. The figures suggest the following.
\begin{itemize}
    \item The physical force method is very sensitive to the values of $N$ in the simulations. The simulated stampede probabilities reach around 0.45 when $N = 40$.
    \item The position errors affect the simulated stampede probabilities of the physical force method, especially when $N = 20$.
    \item The confidence intervals are narrow, which indicates that the expected stampede probabilities would be consistent with repetitive simulations.
\end{itemize}

\subsection{Density method}\label{subsec:density}

\begin{figure}[!ht]
	\centering
	\begin{subfigure}[b]{0.49\columnwidth}
		\includegraphics[width=\columnwidth]{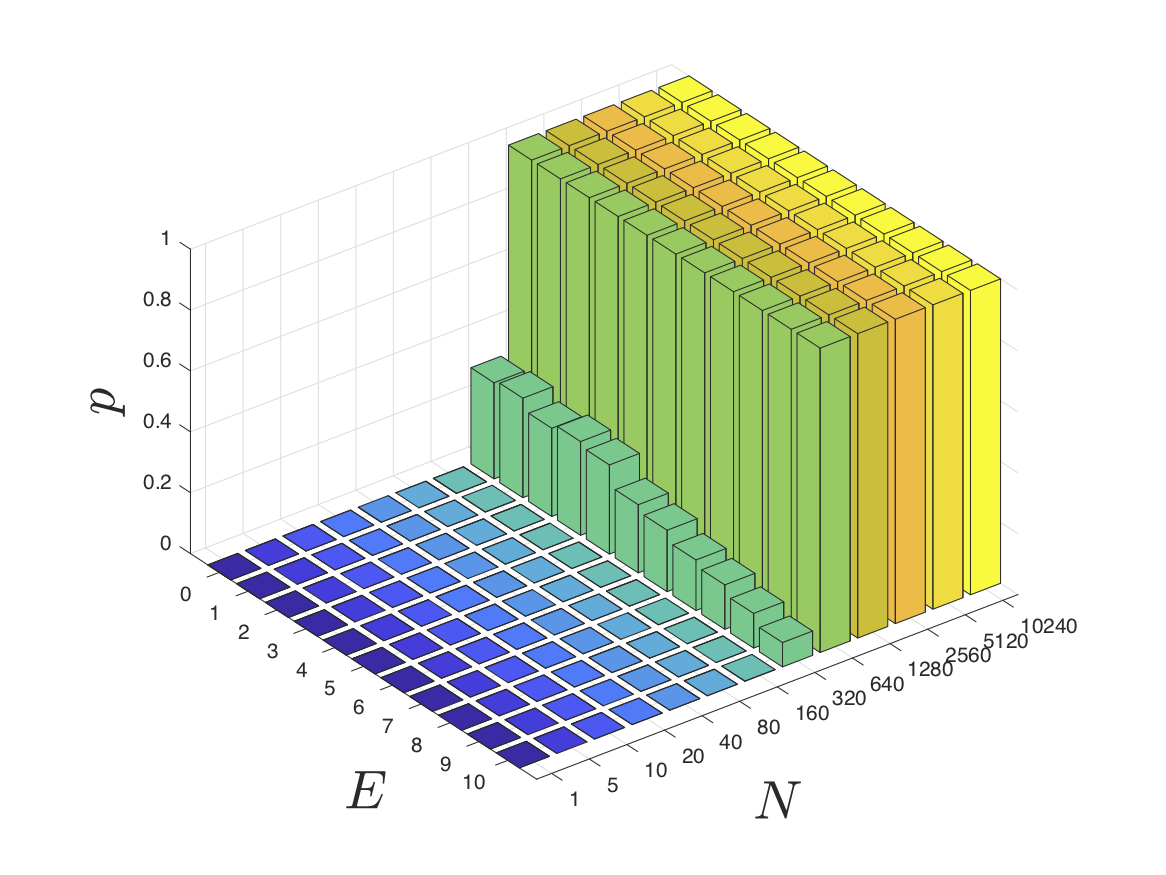}
		\caption{$R = 1$ m}
		\label{fig:density-prob}
	\end{subfigure}
	\begin{subfigure}[b]{0.49\columnwidth}
		\includegraphics[width=\columnwidth]{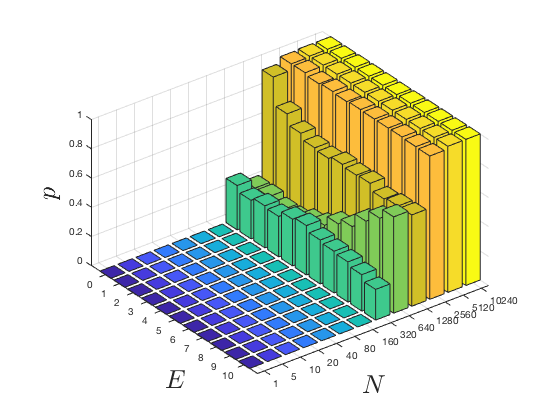}
		\caption{$R = 1.5$ m}
		\label{fig:density-prob-r15}
	\end{subfigure}
	
    \begin{subfigure}[b]{0.49\columnwidth}
		\includegraphics[width=\columnwidth]{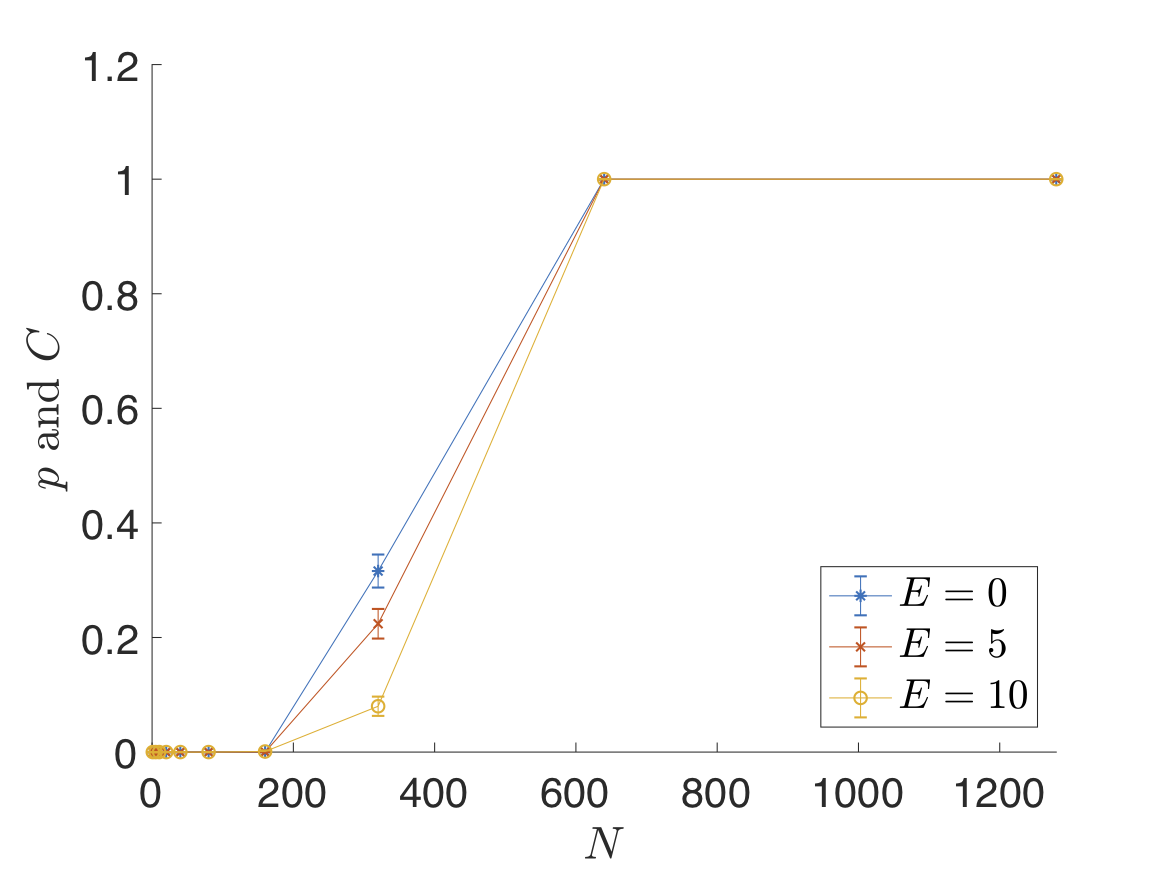}
		\caption{$R = 1$ m}
		\label{fig:density-ci}
	\end{subfigure}
	\begin{subfigure}[b]{0.49\columnwidth}
		\includegraphics[width=\columnwidth]{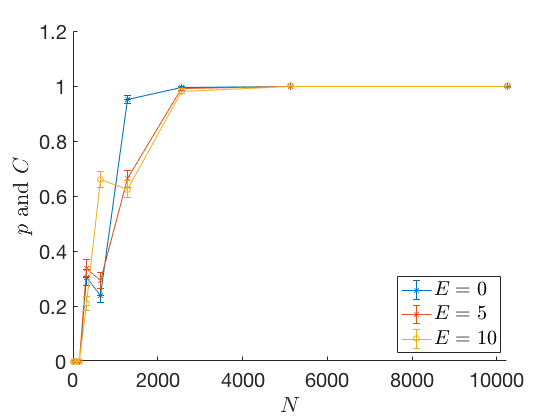}
		\caption{$R = 1.5$ m}
		\label{fig:density-ci-r15}
	\end{subfigure}
	\caption{Simulated stampede probabilities and the corresponding confidence levels in the density method. (a) Simulated stampede probabilities of the density method with $R=1$ m in three-dimensional bar graph of $(N,E,p)$-tuples. (b) Simulated stampede probabilities of the density method with $R=1.5$ m in three-dimensional bar graph. (c) Values of $p$ and $C$ of the density method with $R=1$ m in two-dimensional line plot. (d) Values of $p$ and $C$ of the density method with $R=1.5$ m in two-dimensional line plot.}
	\label{fig:density}
\end{figure}

Results for estimation of $p$ using the density method are given in Fig.~\ref{fig:density}. We vary the value of $R$, setting it to $1$ m and $1.5$ m to adjust the method's sensitivity (as was discussed in Section~\ref{subsec:pressure}). The threshold of density is set to about $7$ m$^{-2}$. 
Examining Fig.~\ref{fig:density} suggests the following. 
\begin{itemize}
    \item The larger the value of $R$---the less sensitive the density method. 
	\item Similar to the results in Fig.~\ref{fig:pressure-prob}, the position errors may affect the value of $p$. For example, Figure~\ref{fig:density-prob} suggests that the number of agents $N$ is equal to 320, the $p$ value varies from $8\%$ to $32.7\%$ for different values of $E$.
	\item As for the other models, the confidence intervals are narrow, indicating high precision and reproducibility of the simulations.
\end{itemize}

% Comparison
\subsection{Comparison}\label{subsec:discussion2}
Comparing all three stampede assessment methods, we can arrive at the following conclusions.
\begin{itemize}
	\item All the three stampede assessment methods can be largely affected by changes to their corresponding thresholds, i.e., $p = 0.04$ s$^{-2}$ for the crowd pressure method, $N = 4500$ N for the physical force method, and $D \approx 7$ m$^{-2}$ for the density method. Moreover, the results of the crowd pressure method and the density method can be affected by changing to the value of $R$.
	\item The simulated stampede probabilities $p$ can be affected by position errors (especially in the crowd pressure method and the density method) if the number of agents is not very small or very large. In the former case $p=0$ and in the latter $p=1$, almost certainly.
	\item In general, the value of $p$ increases with the number of agents.
	\item In all the experiments, the confidence intervals are narrow, suggesting high precision and reproducibility of the simulations.
\end{itemize}

To sum up, the results of the stampede prediction can be affected by the position errors $E$, as well as some input parameters for stampede assessments, e.g., $R$. Thus,  these factors need to be taken into account when implementing stampede measuring techniques. A practitioner can calibrate and validate the model with the empirical data gathered from the agents.

\section{Conclusions and future work}\label{sec:conclusions}
To conclude, in this work we analyze the impact of position errors on the accuracy and precision of stampede prediction. To achieve this goal, we propose an automatic real-time method, called ARES, for stampede prediction. We implement a prototype of ARES using Menge simulation framework. Then, we add position errors to three different stampede assessment methods. We compute the corresponding probabilities of the simulated stampede and compare those results. Finally, we analyze the impact of position errors. Experimental results show that the position errors can change the probability of simulated stampede significantly, even when noise reduction technique, such as the standard Kalman filter, is applied. This implies that analysis of the impact of position errors on crowd simulation is crucial, especially when the GPS signal interference is pronounced. Thus, we argue that more research is needed to take position errors into account when dealing with stampede assessments. The results in this paper can be seen as a starting point to control the position errors in stampede assessments. Future research will focus on exploring potential approaches for mitigating position errors as well as exploring the distributions of position errors from hybrid sources of data (e.g., video cameras and GPS) into this framework, to improve the accuracy of the solution for stampede prediction.

\section*{Acknowledgements}
The work reported in this paper is supported and funded by Ontario Centres of Excellence, Natural Sciences and Engineering Research Council of Canada, and Laipac Technology Inc. We would like to thank Sheik Hoque for performing initial tests on Menge. Last but not least, we would also like to thank the editor and the anonymous reviewers for their valuable suggestions and comments that improved the quality of this paper.

\section*{Availability of data}
Datasets of experimental results are available on Mendeley Data~\cite{dataset18} via \url{http://dx.doi.org/10.17632/tz7gmrzffh.1}

% Noise canceling
\appendix
\section{Noise reduction using Kalman filter}\label{sec:canceling}
In order to test if the Kalman filter could improve the tracking accuracy, we apply a standard Kalman filter to our simulations. The venue map is the same as shown in Fig.~\ref{fig:menge}, and the pedestrian model is the same as described in Section~\ref{subsec:simulation}. We compare filtered estimates to noisy positions (which correspond to measured positions of GPS devices) and true positions. In our case, the state vector of the Kalman filter is $(x_{i,s}, y_{i,s}, v^x_{i,s}, v^y_{i,s})$, where $(x_{i,s}, y_{i,s})$ is the coordinate of agent $i$ at time $s$, and $(v^x_{i,s}, v^y_{i,s})$ is the velocity vector of agent $i$ at time $s$. 

We explore the performance of the Kalman filter in the analysis of a set of agents in \ref{sec:kf_grp} followed by the analysis of two individual agents in \ref{sec:kf_ind}. We summarize our findings in \ref{sec:summary}.

\begin{algorithm}[!ht] % [H] avoids the environment to float
	\caption{Pseudocode to estimate performance of the standard Kalman Filter}\label{alg:kf_experiment}
	\begin{algorithmic}[1] %[1] put number everyline
        \Require $N, S, E$
		\Ensure MAE, KDE
		\State Simulate and store true positions for each agent at every step, while $N = 10240$ and $S = 150$
		\For{$E = 0, 1, \ldots, 10$}
		    \For{each agent}
		        \State Add noise to the position at each step
		        \State Feed these noisy positions to the Kalman filter and estimate the coordinate at the last step
		        \State Compute the error of the noisy (measured) position and the error of the estimated position by comparing to the true position
		    \EndFor
		    \State Compute the MAEs and plot the KDEs
		\EndFor
\end{algorithmic}
\end{algorithm}

\subsection{Estimation of positions for a set of agents}\label{sec:kf_grp}

To explore behaviour of a set of agents, we perform the following set of experiments summarized by the pseudocode in Algorithm~\ref{alg:kf_experiment}. Details of the experiments are given below.

We set the number of agents to $N = 10240$ and the simulation time horizon, deemed $S$, to $S = 150$ s. Each simulation step takes 0.1 s, which means that each agent has 1500 simulation steps. For each agent, we collect and save the position and velocity in each simulation step. We choose 1500 simulation steps, because 1) it provides sufficient prior observations for the Kalman filter, and 2) it spreads agents over the bridge in the end of the simulation (some agents have already left the bridge while others have just reached the first pillar). 

Once the simulation is complete, we perform ten experiments (described below), repeated for ten values of $E = 1,2,\ldots,10$ m. For each value of $E$, we take the actual positions of the agents and add the noise to the coordinates. The noise is drawn from the Rayleigh distribution (see Eq.~(\ref{eq:cdf2}), where the amount of noise is controlled by the value of $E$. The velocities are kept accurate. We then apply the standard Kalman filter to the noisy trajectories (and true velocities) of each agent to estimate the final position of each agent at the last step (i.e., when $S=150$ s). 

After that, for each agent, we compute the absolute errors 1) between the actual and the noisy position at the last step, deemed \textit{measured error} and 2) between the actual and the estimated (by the Kalman filter) position at the last step, deemed \textit{estimated error}. 

% mae table
\begin{table}[!ht]
\centering
%\resizebox{\columnwidth}{!}{
\begin{tabular}{|r|r|r|}
\hline
$E$ & Estimated error & Measured error \\
\hline
1 & 2.68 & 0.89 \\
\hline
2 & 2.71 & 1.76 \\
\hline
3 & 2.74 & 2.66 \\
\hline
4 & 2.77 & 3.59 \\
\hline
5 & 2.80 & 4.44 \\
\hline
6 & 2.83 & 5.30 \\
\hline
7 & 2.86 & 6.23 \\
\hline
8 & 2.90 & 7.19 \\
\hline
9 & 2.93 & 8.00 \\
\hline
10 & 2.96 & 8.83 \\
\hline
\end{tabular}
%}
\caption{MAEs of measured and estimated errors for $E = 1, 2, \ldots, 10$ m. The MAEs are measured in metres. }
\label{tab:mae}
\end{table}

% kernel density estimation
\begin{figure}[!ht]
	\centering
	\begin{subfigure}[b]{0.49\columnwidth}
		\includegraphics[width=\columnwidth]{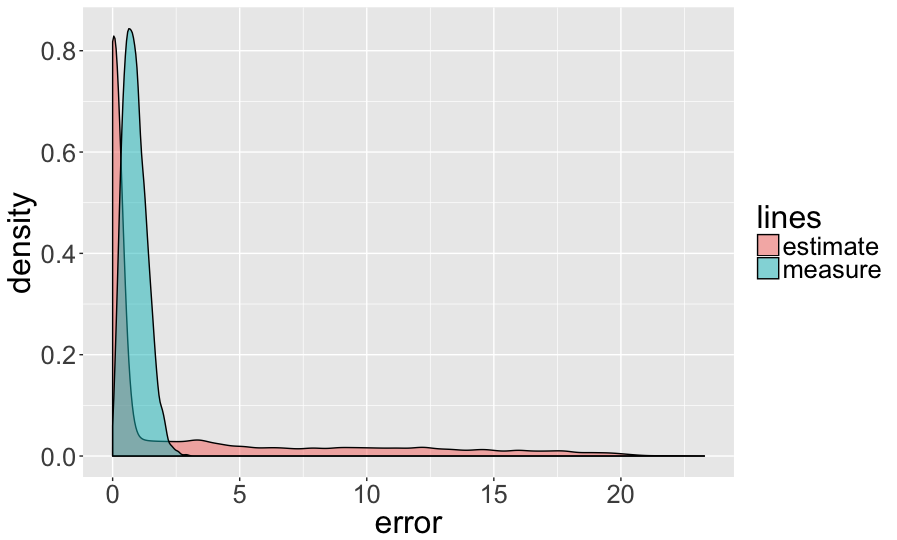}
		\caption{$E = 1$ m}
		\label{fig:d1}
	\end{subfigure}
	\begin{subfigure}[b]{0.49\columnwidth}
		\includegraphics[width=\columnwidth]{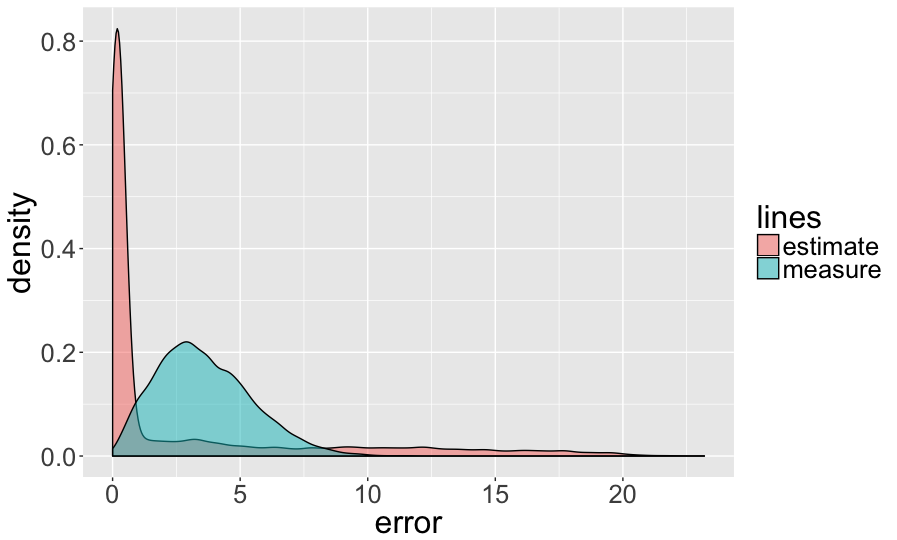}
		\caption{$E = 4$ m}
		\label{fig:d4}
	\end{subfigure}
	
    \begin{subfigure}[b]{0.49\columnwidth}
		\includegraphics[width=\columnwidth]{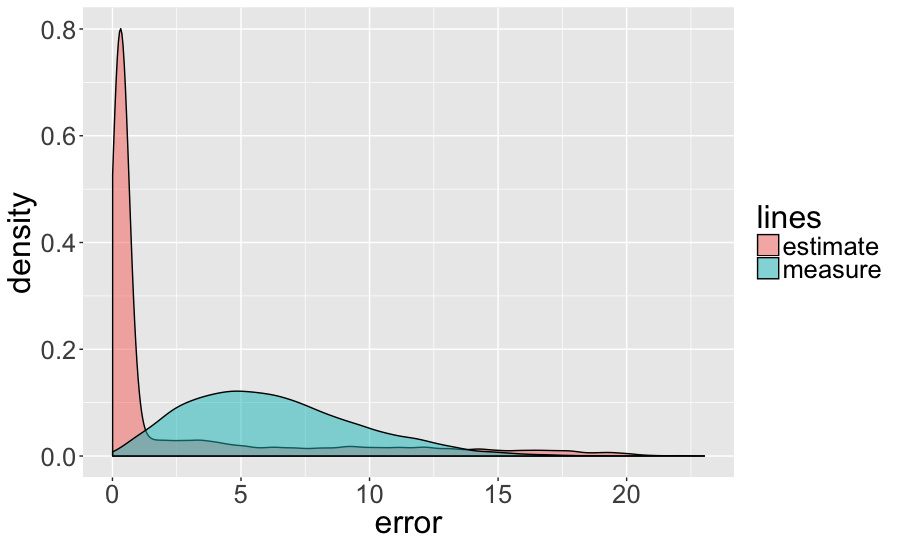}
		\caption{$E = 7$ m}
		\label{fig:d7}
	\end{subfigure}
	\begin{subfigure}[b]{0.49\columnwidth}
		\includegraphics[width=\columnwidth]{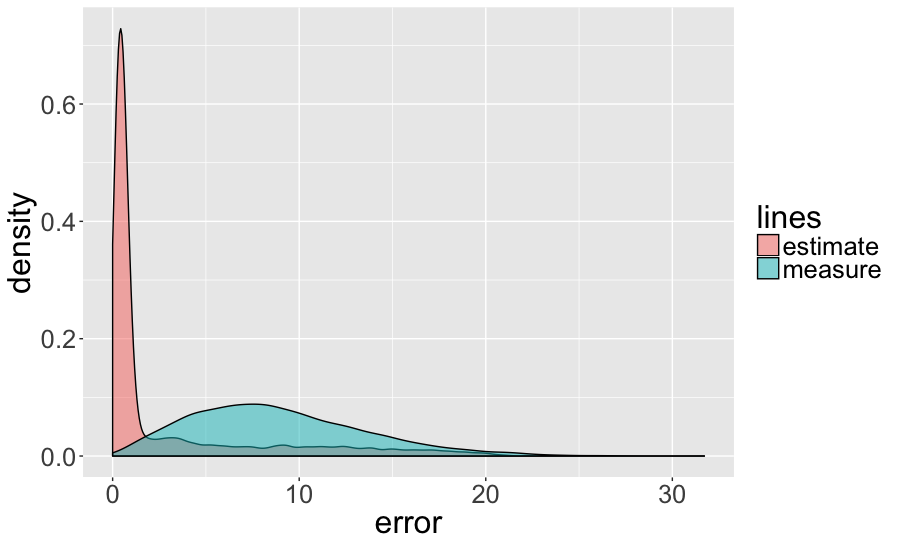}
		\caption{$E = 10$ m}
		\label{fig:d10}
	\end{subfigure}
	\caption{KDEs with $E =$ 1, 4, 7, and 10 m. The x-axis represents the distance errors in metres, and the y-axis represents the probability density. The rose shaded area represents the KDEs of the Kalman filter estimates, and the mint shaded area represents the KDEs of the noisy positions.}
	\label{fig:kernel}
\end{figure}

Finally, we compute the mean absolute errors (MAEs), shown in Table~\ref{tab:mae}. We also plot kernel density estimates (KDEs) of the measured and estimated errors. To preserve space, in Figure~\ref{fig:kernel}, we show KDEs for four representative values of $E$, namely, $E \in \{1, 4, 7, 10\}$.

To compute MAEs, we calculate Euclidean distance errors for each agent's final coordinates and then compute the average of the Euclidean distances of all the agents. Table~\ref{tab:mae} shows the MAEs estimated by the Kalman filter as well as the amount of error added to the actual coordinates. The Kalman filter yields an error, ranging, on average, between 2.6 and 3.0 m. The amount of the estimated error increases monotonically with the growth of $E$. As was shown in Figs.~\ref{fig:pressure-prob},~\ref{fig:force}, and~\ref{fig:density}, the stampede probabilities will be affected by this amount of noise. 

The measured error is smaller than the one yielded by the Kalman filter when $E < 4$ m. Essentially, one should not apply the Kalman filter when $E < 4$ m, as it will decrease the accuracy of the estimated coordinates. It may be beneficial to use the filter when $E \ge 4$ m. However, an error of $\approx 3$ m may lead to significant variation in the positions of the agents relative to each other, reducing the performance of the stampede prediction models. Moreover, KDEs plotted in Figure~\ref{fig:kernel} exhibit heavy right tails: while an average error is  $\approx 3$ m, some agents will have a position error greater than $20$ m.

\subsection{Estimation of positions for two individual agents}\label{sec:kf_ind}

\begin{figure}[!ht]
	\centering
	\begin{subfigure}[b]{\columnwidth}
		\includegraphics[width=0.9\columnwidth]{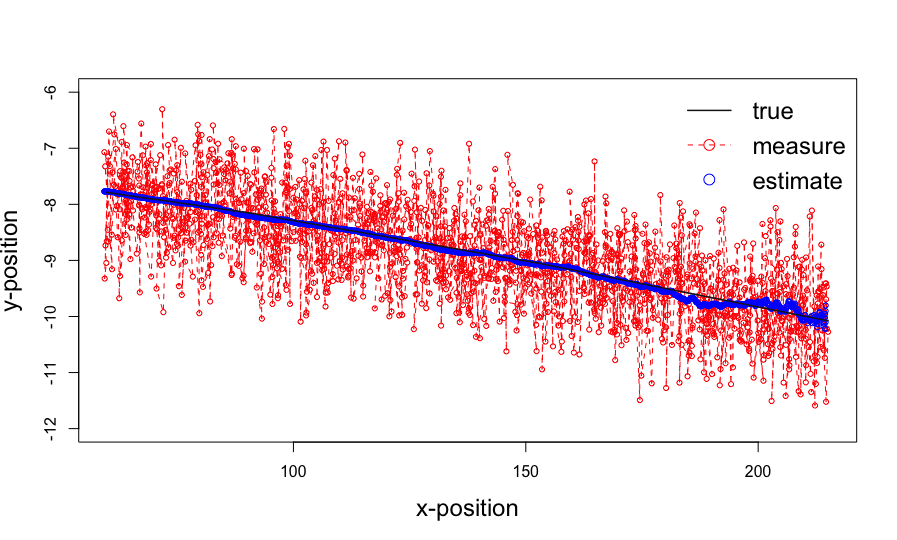}
		\caption{An agent is slowly approaching the first pillar.}
		\label{fig:agent-good}
	\end{subfigure}
	
	\begin{subfigure}[b]{\columnwidth}
		\includegraphics[width=0.9\columnwidth]{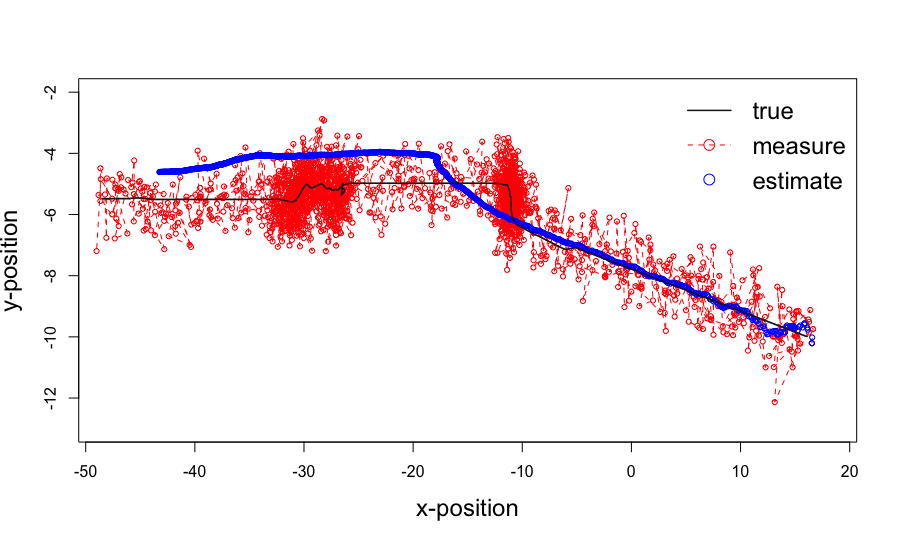}
		\caption{An agent stopped at the first two pillars, and is approaching the third one.}
		\label{fig:agent-bad}
	\end{subfigure}
	\caption{Examples of Kalman filter performance when $E=1$ m. One coordinate unit is equal to one metre.}
	\label{fig:agent-1}
\end{figure}

Why the Kalman filter cannot reduce the MAE below $\approx 3$ m (as shown in Table~\ref{tab:mae})? It has been observed in the past that the Kalman filter has difficulty detecting non-monotonic trajectories~\cite{gustafsson2000adaptive}. We conjecture that the filter faces this challenge in our experiments too. Too support this conjecture, let us take a look at the trajectories of two agents from 10240 agents that we simulated in \ref{sec:kf_grp}.  We then add noise to the trajectories of the agents using Eq.~(\ref{eq:cdf2}) and the setting $E = 1$ m. These trajectories are shown in Fig.~\ref{fig:agent-1}.

In Fig.~\ref{fig:agent-good}, the agent is slowly approaching  the first pillar (moving from right to left), because a lot of agents (which are in front of this agent) block the agent's path. We can see that the agent's trajectory is monotone. In addition, the trajectory estimated by the Kalman filter and the true trajectory starts overlapping from the coordinate around (180, -9.5).

However, the Kalman filter fails to detect abrupt state changes in agent's trajectory; Figure~\ref{fig:agent-bad} depicts such a case. The agent (while moving from right to left) stops around the first two pillars near the coordinates (-10, -6) and (-30, -6) (changing the velocity). The filter has difficulty detecting this change and predicts that the agent keeps moving in a straight line. The agent also  changes the direction near the first pillar, which the filter also does not detect. Thus, the estimates of the Kalman filter have noticeable distances from the true trajectory. 

\subsection{Summary}\label{sec:summary}
To conclude, the standard Kalman filter may help to improve the positional accuracy of GPS when the position errors are high ($E \ge 4$ m). However, the stampede probabilities will still be affected by the amount of position error ($\approx 3$ m) retained by the filter. This level of noise may be explained by the difficulty the filter exhibits while dealing with changing movement patterns. To address this limitation, possible solutions have been proposed in the literature by extending the standard Kalman filter~\cite{moussakhani2014change} or integrating other techniques, e.g., mean shift algorithm~\cite{ying2010intelligent}. We leave the investigation of the applicability of such solutions to future researchers.

\section{The PedVO model}\label{sec:pedvo}

\begin{table}[!ht]
\centering
\resizebox{\columnwidth}{!}{
\begin{tabular}{|l|l|r|}
\hline
Parameter & Description & Value \\
\hline
factor & The factor capturing the relationship between speed and stride length & 1.57 \\
\hline
buffer & The psychological stride buffer required beyond that needed for stride length & 0.9 \\
\hline
tau & The default time horizon for predicting inter-agent collisions & 3.0 \\
\hline
tauObst & The default time horizon for predicting agent-obstacle collisions & 0.1 \\
\hline
turningBias & The default turn bias & 1.0 \\
\hline
density\_aware & Controls if the preferred speed is adjusted to density (true) or not (false) & false \\
\hline
\end{tabular}
}
\caption{Default parameters of PedVO adopted in our experiments.}
\label{tab:pedvo}
\end{table}

As mentioned in Section~\ref{subsec:simulation}, we use the fundamental diagram as the velocity modifier, which represents the relationship between speed and density. The pedestrian model adopted, namely, PedVO~\cite{curtis2014pedestrian}, also captures the underlying physiological and psychological factors besides the fundamental diagram. In our experiments, we use the default settings of PedVO in the fundamental diagram example provided by Menge. These settings are shown in Table~\ref{tab:pedvo}.

\bibliographystyle{abbrv}

\begin{thebibliography}{10}
	
	\bibitem{online:ebracelet}
	{\relax Al Arabiya News}.
	\newblock {S}audi {A}rabia may soon issue e-bracelets for all hajj pilgrims,
	2015.
	\newblock [Online; accessed 1-July-2018].
	
	\bibitem{almeida2011crowd}
	J.~E. Almeida, R.~Rosseti, and A.~L. Coelho.
	\newblock Crowd simulation modeling applied to emergency and evacuation
	simulations using multi-agent systems.
	\newblock In {\em Proceedings of the 6th Doctoral Symposium on Informatics
		Engineering}, pages 93--104. FEUP, 2011.
	
	\bibitem{alonso2014simulation}
	F.~Alonso-Marroquin, J.~Busch, C.~Chiew, C.~Lozano, and
	{\'A}.~Ram{\'\i}rez-G{\'o}mez.
	\newblock Simulation of counterflow pedestrian dynamics using spheropolygons.
	\newblock {\em Physical Review E}, 90(6):063305, 2014.
	
	\bibitem{online:shanghai}
	{\relax BBC News}.
	\newblock Shanghai new year crush kills 36, 2015.
	\newblock [Online; accessed 1-July-2018].
	
	\bibitem{bostanci2018sensor}
	E.~Bostanci, B.~Bostanci, N.~Kanwal, and A.~F. Clark.
	\newblock Sensor fusion of camera, {GPS} and {IMU} using fuzzy adaptive
	multiple motion models.
	\newblock {\em Soft Computing}, 22(8):2619--2632, 2018.
	
	\bibitem{brown1992introduction}
	R.~G. Brown, P.~Y. Hwang, et~al.
	\newblock {\em Introduction to random signals and applied {K}alman filtering},
	volume~3.
	\newblock Wiley New York, 1992.
	
	\bibitem{browning2006effects}
	R.~C. Browning, E.~A. Baker, J.~A. Herron, and R.~Kram.
	\newblock Effects of obesity and sex on the energetic cost and preferred speed
	of walking.
	\newblock {\em Journal of Applied Physiology}, 100(2):390--398, 2006.
	
	\bibitem{callisaya2017cognitive}
	M.~L. Callisaya, C.~P. Launay, V.~K. Srikanth, J.~Verghese, G.~Allali, and
	O.~Beauchet.
	\newblock Cognitive status, fast walking speed and walking speed reserve-the
	gait and alzheimer interactions tracking ({GAIT}) study.
	\newblock {\em Geroscience}, 39(2):231--239, 2017.
	
	\bibitem{corbetta2016continuous}
	A.~Corbetta, J.~Meeusen, C.~Lee, and F.~Toschi.
	\newblock Continuous measurements of real-life bidirectional pedestrian flows
	on a wide walkway.
	\newblock In {\em Proceedings of Pedestrian and Evacuation Dynamics 2016},
	pages 18--24, 2016.
	
	\bibitem{curtis2016menge}
	S.~Curtis, A.~Best, and D.~Manocha.
	\newblock Menge: A modular framework for simulating crowd movement.
	\newblock {\em Collective Dynamics}, 1:1--40, 2016.
	
	\bibitem{curtis2014pedestrian}
	S.~Curtis and D.~Manocha.
	\newblock Pedestrian simulation using geometric reasoning in velocity space.
	\newblock In U.~Weidmann, U.~Kirsch, and M.~Schreckenberg, editors, {\em
		Proceedings of Pedestrian and Evacuation Dynamics 2012}, pages 875--890,
	Switzerland, 2014. Springer.
	
	\bibitem{dridi2015list}
	M.~H. Dridi.
	\newblock List parameters influencing the pedestrian movement and pedestrian
	database.
	\newblock {\em International Journal of Social Science Studies}, 3(4):94--106,
	2015.
	
	\bibitem{eliasson2014kalman}
	M.~Eliasson.
	\newblock A {K}alman filter approach to reduce position error for pedestrian
	applications in areas of bad {GPS} reception, 2014.
	
	\bibitem{elliott1993football}
	D.~Elliott and D.~Smith.
	\newblock Football stadia disasters in the {U}nited {K}ingdom: learning from
	tragedy?
	\newblock {\em Industrial \& Environmental Crisis Quarterly}, 7(3):205--229,
	1993.
	
	\bibitem{franke2015smart}
	T.~Franke, P.~Lukowicz, and U.~Blanke.
	\newblock Smart crowds in smart cities: Real life, city scale deployments of a
	smartphone based participatory crowd management platform.
	\newblock {\em Journal of Internet Services and Applications}, 6(1):27, 2015.
	
	\bibitem{garnett2015comparison}
	R.~Garnett and R.~Stewart.
	\newblock Comparison of {GPS} units and mobile {A}pple {GPS} capabilities in an
	urban landscape.
	\newblock {\em Cartography and Geographic Information Science}, 42(1):1--8,
	2015.
	
	\bibitem{groves2014four}
	P.~D. Groves, L.~Wang, D.~Walter, H.~Martin, K.~Voutsis, and Z.~Jiang.
	\newblock The four key challenges of advanced multisensor navigation and
	positioning.
	\newblock In {\em Proceedings of 2014 IEEE/ION Position, Location and
		Navigation Symposium (PLANS)}, pages 773--792. IEEE, 2014.
	
	\bibitem{gustafsson2000adaptive}
	F.~Gustafsson and F.~Gustafsson.
	\newblock {\em Adaptive filtering and change detection}, volume~1.
	\newblock Citeseer, 2000.
	
	\bibitem{helbing2002simulation}
	D.~Helbing, I.~J. Farkas, P.~Molnar, and T.~Vicsek.
	\newblock Simulation of pedestrian crowds in normal and evacuation situations.
	\newblock {\em Pedestrian and evacuation dynamics}, 21(2):21--58, 2002.
	
	\bibitem{helbing2007dynamics}
	D.~Helbing, A.~Johansson, and H.~Z. Al-Abideen.
	\newblock Dynamics of crowd disasters: An empirical study.
	\newblock {\em Physical review E}, 75(4), 2007.
	
	\bibitem{helbing2007supplementary}
	D.~Helbing, A.~Johansson, and H.~Z. Al-Abideen.
	\newblock The dynamics of crowd disasters: An empirical study (supplementary
	information), 2007.
	\newblock [Online; accessed 7-February-2018].
	
	\bibitem{johansson2008crowd}
	A.~Johansson, D.~Helbing, H.~Z. Al-Abideen, and S.~Al-Bosta.
	\newblock From crowd dynamics to crowd safety: A video-based analysis.
	\newblock {\em Advances in Complex Systems}, 11(04):497--527, 2008.
	
	\bibitem{johansson2007specification}
	A.~Johansson, D.~Helbing, and P.~K. Shukla.
	\newblock Specification of the social force pedestrian model by evolutionary
	adjustment to video tracking data.
	\newblock {\em Advances in complex systems}, 10(supp02):271--288, 2007.
	
	\bibitem{junior2010crowd}
	J.~C. S.~J. Junior, S.~R. Musse, and C.~R. Jung.
	\newblock Crowd analysis using computer vision techniques.
	\newblock {\em IEEE Signal Processing Magazine}, 27(5):66--77, 2010.
	
	\bibitem{krausz2012loveparade}
	B.~Krausz and C.~Bauckhage.
	\newblock Loveparade 2010: Automatic video analysis of a crowd disaster.
	\newblock {\em Computer Vision and Image Understanding}, 116(3):307--319, 2012.
	
	\bibitem{kurilkin2016comparison}
	A.~V. Kurilkin and S.~V. Ivanov.
	\newblock A comparison of methods to detect people flow using video processing.
	\newblock {\em Procedia Computer Science}, 101:125--134, 2016.
	
	\bibitem{launer1996weight}
	L.~J. Launer and T.~Harris.
	\newblock Weight, height and body mass index distributions in geographically
	and ethnically diverse samples of older persons.
	\newblock {\em Age and Ageing}, 25(4):300--306, 1996.
	
	\bibitem{lee2005exploring}
	R.~S. Lee and R.~L. Hughes.
	\newblock Exploring trampling and crushing in a crowd.
	\newblock {\em Journal of transportation engineering}, 131(8):575--582, 2005.
	
	\bibitem{lee2006prediction}
	R.~S. Lee and R.~L. Hughes.
	\newblock Prediction of human crowd pressures.
	\newblock {\em Accident analysis \& prevention}, 38(4):712--722, 2006.
	
	\bibitem{li2000gps}
	J.~Li, K.~Miyashita, T.~Kato, and S.~Miyazaki.
	\newblock {GPS} time series modeling by autoregressive moving average method:
	Application to the crustal deformation in central {J}apan.
	\newblock {\em Earth, planets and space}, 52(3):155--162, 2000.
	
	\bibitem{ying2010intelligent}
	Y.~Li, Y.~Pang, Z.~Li, and Y.~Liu.
	\newblock An intelligent tracking technology based on {K}alman and mean shift
	algorithm.
	\newblock In {\em Proceedings of the 2nd International Conference on Computer
		Modeling and Simulation (ICCMS)}, volume~1, pages 107--109. IEEE, 2010.
	
	\bibitem{littlefield2012metric}
	D.~Littlefield.
	\newblock {\em Metric handbook: Planning and design data}.
	\newblock Routledge, 2012.
	
	\bibitem{liu2012energy}
	J.~Liu, B.~Priyantha, T.~Hart, H.~S. Ramos, A.~A. Loureiro, and Q.~Wang.
	\newblock Energy efficient {GPS} sensing with cloud offloading.
	\newblock In {\em Proceedings of the 10th ACM Conference on Embedded Network
		Sensor Systems}, pages 85--98. ACM, 2012.
	
	\bibitem{mahmood2017analyzing}
	I.~Mahmood, M.~Haris, and H.~Sarjoughian.
	\newblock Analyzing emergency evacuation strategies for mass gatherings using
	crowd simulation and analysis framework: Hajj scenario.
	\newblock In {\em Proceedings of the 2017 ACM SIGSIM Conference on Principles
		of Advanced Discrete Simulation}, pages 231--240. ACM, 2017.
	
	\bibitem{marana1999estimating}
	A.~N. Marana, L.~D.~F. Costa, R.~Lotufo, and S.~A. Velastin.
	\newblock Estimating crowd density with {M}inkowski fractal dimension.
	\newblock In {\em Proceedings of 1999 IEEE International Conference on
		Acoustics, Speech, and Signal Processing}, pages 3521--3524. IEEE, 1999.
	
	\bibitem{mousavi2015analyzing}
	H.~Mousavi, S.~Mohammadi, A.~Perina, R.~Chellali, and V.~Mur.
	\newblock Analyzing tracklets for the detection of abnormal crowd behavior.
	\newblock In {\em Proceedings of 2015 IEEE Winter Conference on Applications of
		Computer Vision (WACV)}, pages 148--155. IEEE, 2015.
	
	\bibitem{moussakhani2014change}
	B.~Moussakhani, J.~T. Fl{\aa}m, T.~A. Ramstad, and I.~Balasingham.
	\newblock On change detection in a {K}alman filter based tracking problem.
	\newblock {\em Signal processing}, 105:268--276, 2014.
	
	\bibitem{online:smartphone_penetration}
	Newzoo.
	\newblock Top countries/markets by smartphone penetration \& users, 2018.
	\newblock [Online; accessed 15-August-2018].
	
	\bibitem{online:gps}
	{\relax Official U.S. government information about the Global Positioning
		System (GPS) and related topics}.
	\newblock {GPS} accuracy, 2017.
	\newblock [Online; accessed 15-August-2018].
	
	\bibitem{papoulis2002probability}
	A.~Papoulis and S.~U. Pillai.
	\newblock {\em Probability, random variables, and stochastic processes}.
	\newblock Tata McGraw-Hill Education, New York, 4th edition, 2002.
	
	\bibitem{petovello2015does}
	M.~PETOVELLO.
	\newblock How does a {GNSS} receiver estimate velocity?
	\newblock {\em Inside GNSS}, pages 38--41, 2015.
	
	\bibitem{sabokrou2017deep}
	M.~Sabokrou, M.~Fayyaz, M.~Fathy, and R.~Klette.
	\newblock Deep--cascade: Cascading 3{D} deep neural networks for fast anomaly
	detection and localization in crowded scenes.
	\newblock {\em IEEE Transactions on Image Processing}, 26(4):1992--2004, 2017.
	
	\bibitem{sabokrou2018deep}
	M.~Sabokrou, M.~Fayyaz, M.~Fathy, Z.~Moayed, and R.~Klette.
	\newblock Deep-anomaly: Fully convolutional neural network for fast anomaly
	detection in crowded scenes.
	\newblock {\em Computer Vision and Image Understanding}, 2018.
	
	\bibitem{sharma2016review}
	D.~Sharma, A.~P. Bhondekar, A.~Shukla, and C.~Ghanshyam.
	\newblock A review on technological advancements in crowd management.
	\newblock {\em Journal of Ambient Intelligence and Humanized Computing},
	9(3):1--11, 2016.
	
	\bibitem{online:smartphone_worldwide}
	Statista.
	\newblock Number of smartphone users worldwide 2014 - 2020, 2016.
	\newblock [Online; accessed 15-August-2018].
	
	\bibitem{steffen2010methods}
	B.~Steffen and A.~Seyfried.
	\newblock Methods for measuring pedestrian density, flow, speed and direction
	with minimal scatter.
	\newblock {\em Physica A: Statistical mechanics and its applications},
	389(9):1902--1910, 2010.
	
	\bibitem{stojanovic2014indoor}
	D.~Stojanovi{\'c} and N.~Stojanovi{\'c}.
	\newblock Indoor localization and tracking: Methods, technologies and research
	challenges.
	\newblock {\em Facta Universitatis, Series: Automatic Control and Robotics},
	13(1):57--72, 2014.
	
	\bibitem{suzuki2011high}
	T.~Suzuki, M.~Kitamura, Y.~Amano, and T.~Hashizume.
	\newblock High-accuracy {GPS} and {GLONASS} positioning by multipath mitigation
	using omnidirectional infrared camera.
	\newblock In {\em Proceedings of 2011 IEEE International Conference on Robotics
		and Automation (ICRA)}, pages 311--316. IEEE, 2011.
	
	\bibitem{tange2018gnu}
	O.~Tange.
	\newblock {GNU} parallel 2018, 2018.
	
	\bibitem{online:hajj}
	{\relax The Express Tribune}.
	\newblock Hajj stampede death toll rises to 2,177, 2015.
	\newblock [Online; accessed 1-July-2018].
	
	\bibitem{online:mumbai}
	{\relax The New York Times}.
	\newblock Stampede at {M}umbai railway station kills at least 22, 2017.
	\newblock [Online; accessed 1-July-2018].
	
	\bibitem{online:turin}
	{\relax The Telegraph News}.
	\newblock More than 1,500 {J}uventus fans injured in stampede in {T}urin, 2017.
	\newblock [Online; accessed 1-July-2018].
	
	\bibitem{tordeux2015quantitative}
	A.~Tordeux, J.~Zhang, B.~Steffen, and A.~Seyfried.
	\newblock Quantitative comparison of estimations for the density within
	pedestrian streams.
	\newblock {\em Journal of Statistical Mechanics: Theory and Experiment},
	2015(6):P06030, 2015.
	
	\bibitem{online:ucsd}
	UCSD.
	\newblock Anomaly detection dataset, 2016.
	\newblock [Online; accessed 15-August-2018].
	
	\bibitem{van2011reciprocal}
	J.~Van Den~Berg, S.~J. Guy, M.~Lin, and D.~Manocha.
	\newblock Reciprocal n-body collision avoidance.
	\newblock In {\em Robotics research}, pages 3--19. Springer, 2011.
	
	\bibitem{van2009gps}
	F.~S.~T. Van~Diggelen.
	\newblock {\em {A-GPS}: {A}ssisted {GPS}, {GNSS}, and {SBAS}}.
	\newblock Artech House, 2009.
	
	\bibitem{walpole2012weight}
	S.~C. Walpole, D.~Prieto-Merino, P.~Edwards, J.~Cleland, G.~Stevens, and
	I.~Roberts.
	\newblock The weight of nations: An estimation of adult human biomass.
	\newblock {\em BMC public health}, 12(1):439, 2012.
	
	\bibitem{wang2008error}
	F.~Wang, X.~Zhang, and J.~Huang.
	\newblock Error analysis and accuracy assessment of {GPS} absolute velocity
	determination without {SA}.
	\newblock {\em Geo-spatial Information Science}, 11(2):133--138, 2008.
	
	\bibitem{wilson2002gps}
	D.~L. Wilson.
	\newblock {GPS} horizontal position accuracy.
	\newblock {\em Global Positioning System Accuracy webpage}, 2002.
	
	\bibitem{wirz2012inferring}
	M.~Wirz, T.~Franke, D.~Roggen, E.~Mitleton-Kelly, P.~Lukowicz, and
	G.~Tr{\"o}ster.
	\newblock Inferring crowd conditions from pedestrians' location traces for
	real-time crowd monitoring during city-scale mass gatherings.
	\newblock In {\em Proceedings of 2012 IEEE 21st International Workshop on
		Enabling Technologies: Infrastructure for Collaborative Enterprises
		(WETICE)}, pages 367--372. IEEE, 2012.
	
	\bibitem{yamaguchi2006gps}
	S.~Yamaguchi and T.~Tanaka.
	\newblock {GPS} standard positioning using {K}alman filter.
	\newblock In {\em Proceedings of 2006 International Joint Conference
		SICE-ICASE}, pages 1351--1354. IEEE, 2006.
	
	\bibitem{yin2013case}
	R.~K. Yin.
	\newblock {\em Case study research: Design and methods}.
	\newblock Sage publications, 5th edition, 2013.
	
	\bibitem{zandbergen2008positional}
	P.~A. Zandbergen.
	\newblock Positional accuracy of spatial data: Non-normal distributions and a
	critique of the national standard for spatial data accuracy.
	\newblock {\em Transactions in GIS}, 12(1):103--130, 2008.
	
	\bibitem{zandbergen2011positional}
	P.~A. Zandbergen and S.~J. Barbeau.
	\newblock Positional accuracy of assisted {GPS} data from high-sensitivity
	{GPS}-enabled mobile phones.
	\newblock {\em The Journal of Navigation}, 64(3):381--399, 2011.
	
	\bibitem{zkebala2012pedestrian}
	J.~Z{\k{e}}bala, P.~Ci{\k{e}}pka, and A.~RezA.
	\newblock Pedestrian acceleration and speeds.
	\newblock {\em Probl. Forensic Sci.}, 91:227--234, 2012.
	
	\bibitem{zelenkov2008accuracy}
	A.~Zelenkov, A.~Kluga, and E.~Grab.
	\newblock Accuracy estimation of {GPS} receiver parameters with re-reference
	system in static mode.
	\newblock {\em Telecommunications \& Electronics}, 8(31):31--36, 2008.
	
	\bibitem{dataset18}
	L.~Zhang, D.~Lai, and A.~Miranskyy.
	\newblock Datasets for the impact of position errors on crowd simulation.
	\newblock {\em Mendeley Data}, 2018.
	
	\bibitem{zhou2017early}
	J.~Zhou, H.~Pei, and H.~Wu.
	\newblock Early warning of human crowds based on query data from {B}aidu maps:
	Analysis based on {S}hanghai stampede.
	\newblock In {\em Big Data Support of Urban Planning and Management}, pages
	19--41. Springer, Cham, 2017.
	
	\bibitem{zhou2016spatial}
	S.~Zhou, W.~Shen, D.~Zeng, M.~Fang, Y.~Wei, and Z.~Zhang.
	\newblock Spatial--temporal convolutional neural networks for anomaly detection
	and localization in crowded scenes.
	\newblock {\em Signal Processing: Image Communication}, 47:358--368, 2016.
	
\end{thebibliography}

\end{document}